\pdfoutput=1
\documentclass[12pt, draftclsnofoot, onecolumn]{IEEEtran}
\usepackage{cite}
\usepackage{indentfirst}
\usepackage{graphicx}
\usepackage{changepage}
\usepackage{stfloats}
\usepackage{amsfonts,amssymb}
\usepackage{bm}
\usepackage{algorithm}
\usepackage{algorithmic}
\usepackage{amsmath}
\usepackage{amsmath,amssymb,amsfonts}
\usepackage{times}
\usepackage{url}
\usepackage{cite}
\usepackage{textcomp}
\usepackage{xcolor}
\usepackage{color}
\usepackage{setspace}
\usepackage{enumitem}
\usepackage{float}
\usepackage{diagbox}
\usepackage[T1]{fontenc}
\usepackage[utf8]{inputenc}
\usepackage{authblk}
\usepackage{threeparttable}
\usepackage{booktabs}
\usepackage{footnote}
\usepackage{graphicx}
\usepackage{pifont}
\usepackage{subfigure}
\usepackage{mathrsfs}
\allowdisplaybreaks[4]

\newenvironment{proof}{{\indent  \it Proof:}}{\hfill $\blacksquare$}

\newtheorem{remark}{\bf Remark}
\newtheorem{thm}{\bf Lemma}
\newtheorem{pro}{\bf Proposition}
\newtheorem{cor}{\bf Corollary}

\newenvironment{shrinkeq}[1]
{ \bgroup
	\addtolength\abovedisplayshortskip{#1}
	\addtolength\abovedisplayskip{#1}
	\addtolength\belowdisplayshortskip{#1}
	\addtolength\belowdisplayskip{#1}}
{\egroup\ignorespacesafterend}

\begin{document}
\title{Multi-UAV Collaborative Sensing and Communication: Joint Task Allocation and Power Optimization}

\author{
	Kaitao Meng, \textit{Member, IEEE}, Xiaofan He, \textit{Senior Member, IEEE}, Qingqing Wu, \textit{Senior Member, IEEE}, and Deshi Li
	\thanks{{K. Meng and Q. Wu are with the State Key Laboratory of Internet of Things for Smart City, University of Macau, Macau, 999078, China. {(emails: \{kaitaomeng, qingqingwu\}@um.edu.mo)}. X. He and D. Li are with the Electronic Information School, Wuhan University, Wuhan, 430072, China. {(emails: \{xiaofanhe, dsli\}@whu.edu.cn)}}. }
}

\maketitle

\vspace{-18mm}
\begin{abstract}
Due to the features of on-demand deployment and flexible observation, unmanned aerial vehicles (UAVs) are promising for serving as the next-generation aerial sensors by using their onboard sensing devices. Compared to a single UAV with limited sensing coverage and communication capability, multi-UAV cooperation is able to realize more effective sensing and transmission (S\&T) services, and delivers the sensory data to the control center more efficiently for further analysis. Nevertheless, most existing works on multi-UAV sensing mainly focus on mutually exclusive task allocation and independent data transmission, which did not fully exploit the benefit of multi-UAV sensing and communication. Motivated by this, we propose a novel multi-UAV cooperative S\&T scheme with overlapped sensing task allocation. Although overlapped task allocation may sound counter-intuitive, it can actually foster cooperative transmission among multiple UAVs through a virtual multi-antenna system and thus reduce the overall sensing mission completion time. To obtain the optimal task allocation and transmit power of the proposed scheme, a mission completion time minimization problem is formulated. To solve this problem, a condition that specifies whether it is necessary for the UAVs to perform overlapped sensing is derived. For the cases of overlapped sensing, this time minimization problem is transformed into a monotonic optimization and is solved by the generic Polyblock algorithm. To efficiently evaluate the mission completion time in each iteration of the Polyblock algorithm, new auxiliary variables are introduced to decouple the otherwise sophisticated joint optimization of transmission time and power. While for the degenerated case of non-overlapped sensing, the closed-form expression of the optimal transmission time is derived, which provides insights into the optimal solution and facilitates the design of an efficient double-loop binary search algorithm to optimally solve the degenerated problem. Finally, simulation results demonstrate that the proposed scheme significantly reduces the mission completion time over benchmark schemes.

\end{abstract}
\vspace{-6mm}

\section{Introduction}
Driven by its high mobility and flexibility, unmanned aerial vehicle (UAV) is expected to play an increasingly important role in future sensing systems to provide larger sensing coverage, more flexible observation, and richer target information \cite{LiuEnergy, Meng2021Space, Hua2019Energy}. In particular, sensor-equipped UAVs can act as data providers to perform sensing tasks in areas that are difficult to reach by terrestrial vehicles, such as disaster rescue, power-line inspection, aerial 3D reconstruction, etc \cite{Liu2019Resource, disaster_Zhang}. Benefited from the line-of-sight (LoS)-dominated channels of cellular-connected UAVs \cite{Zeng2019Accessing}, the collected sensory data can be directly transmitted to the base station (BS) more efficiently via wireless communications, instead of forcing UAVs to fly back to the control centers. In this way, faster mission response can be achieved as compared to the traditional data collection systems \cite{Mozaffari, Meng2022ThroughputMaximization}. Nonetheless, due to its limited sensing and communication capabilities, single-UAV sensing may not perform well, especially when multiple geographically distributed tasks are involved \cite{OpportunitiesChallenges, SenseSendProtocol}. Therefore, there is an urgent need to develop multi-UAV cooperative sensing and transmission (S\&T) mechanisms that can provide more effective data collection and transmission services \cite{Wu2018JointTrajectory}.
\par
In the literature, the idea of using UAV-relay to improve the latency performance in multi-UAV sensing and communication systems has been widely explored \cite{Scherer2020Surveillance, zhang2019cellular, SensingQualityRelay}. For example, a multi-UAV collaboration S\&T scheme was developed in \cite{Scherer2020Surveillance}, where the sensory data was transmitted in a store-and-forward mode along minimum latency paths to satisfy the predefined latency requirements. In \cite{zhang2019cellular}, a cooperative multi-UAV transfer and relay protocol was proposed to maximize the uplink sum-rate by optimizing the channel allocation and UAV flight speed. Moreover, the sensory data collected by one sensing UAV can be compressed and transmitted to the BS through the cooperation of multi-hop relay UAVs, and the compression ratio and UAV location can be jointly optimized to improve the data freshness \cite{SensingQualityRelay}. Although employing UAVs to relay sensory data can reduce communication latency, such dedicated utilization of UAVs as relays will inevitably lead to lower sensing efficiency. On the other hand, relay communication requires each UAV to amplify/decode the received signal from the previous hop before sending it to the next UAV, which unfavorably brings transmission noise and delay \cite{Behnad2015Distributed}.
\par  
Besides relay communication, the cooperative transmission via distributed multi-antenna technique is another effective way to improve the communication capability of multi-UAV systems \cite{meng2022uavMagazine, Wu2021Comprehensive, Jia2018Joint, Kwon2017Effective}. Specifically, when multiple UAVs possess identical sensory data, they can form a virtual multi-antenna system in the sky and transmit cooperatively to improve communication performance. For instance, multiple UAVs equipped with a single antenna can work as a coordinate multipoint (CoMP) system to enhance interference mitigation and maximize the network throughput by exploiting the high mobility of UAVs \cite{Liu2019CoMP}. The authors in \cite{Mozaffari2019Communications} studied a multi-UAV-based antenna array system where the service time is minimized by optimizing the wireless transmission time as well as the movement time of UAVs. Moreover, in \cite{Hanna2019Distributed}, authors proposed a UAV swarm backhaul scheme with high multiplexing gain, where UAV's positions are jointly optimized to enhance the channel gain of the virtual multiple-input and multiple-output (MIMO) backhaul links. Besides, UAV swarm-enabled virtual MIMO communications were investigated with consideration of physical layer security in \cite{Jung2022Security} and practical features of signal propagation in \cite{Qian2022Configurable}, respectively. However, the mechanisms considered in \cite{Jia2018Joint, Kwon2017Effective, Liu2019CoMP, Hanna2019Distributed, Jung2022Security, Mozaffari2019Communications, Qian2022Configurable} generally require explicit data sharing among multiple UAVs before cooperative transmission, causing extra time and energy consumption. 
\par
In multi-UAV cooperative sensing and communication systems, effective task allocation can further reduce mission completion time \cite{Gao2022UAV, Wang2018MultiTask, Gu2018Multiple}. Especially, UAVs with high mobility can provide new degrees of freedom to flexibly assign the sensing tasks and form a dynamic sensor team \cite{guerra2020dynamic}. It is worth noting that the conventional wisdom of sensing task allocation is to prevent UAVs from repeating the same tasks, since it is commonly believed that such task allocation strategies can improve the efficiency of data acquisition and enlarge sensing coverage \cite{Li2022Intelligent, peng2018wide, MobileSensingUAV, Chen2020PerformanceUAV, CompletionMinimizationMultiUAV, CellularUAVDeviceCommunications, Wang2021Constrained}. For example, a UAV swarm-based hierarchical network architecture was developed to jointly schedule sensing, computing, and communication resources, thereby reducing the delay caused by overlapping sensing operations \cite{Li2022Intelligent}. In \cite{peng2018wide}, multiple UAVs and ground vehicles are employed to cooperatively perform inspection tasks in parallel, thereby enhancing efficiency and expanding service areas. In \cite{MobileSensingUAV}, a cooperative crowd sensing mechanism is developed, by employing multiple UAVs to perform tasks in different sub-regions simultaneously, so as to improve energy efficiency. A novel joint sensing and communication scheme was proposed in a multi-UAV cooperation system to simultaneously conduct radar sensing and sensing data transmission, where the sensing mission is allocated to the UAVs without overlapped tasks \cite{Chen2020PerformanceUAV}. In \cite{CompletionMinimizationMultiUAV}, two multi-UAV-enabled collaborative data collection methods were proposed, where each UAV collects data from sensors in the disjoint cluster. In \cite{CellularUAVDeviceCommunications}, the sensory data collected by each UAV is transmitted through the UAV-to-device link or the UAV-BS link independently according to the communication channel gain. However, with these conventional task allocations, the UAVs often hold different sensory data and cannot form a virtual multi-antenna system for efficient collaborative transmission. 
\par

\begin{figure}[t]
	\centering
	\setlength{\abovecaptionskip}{0.cm}
	\includegraphics[width=15cm]{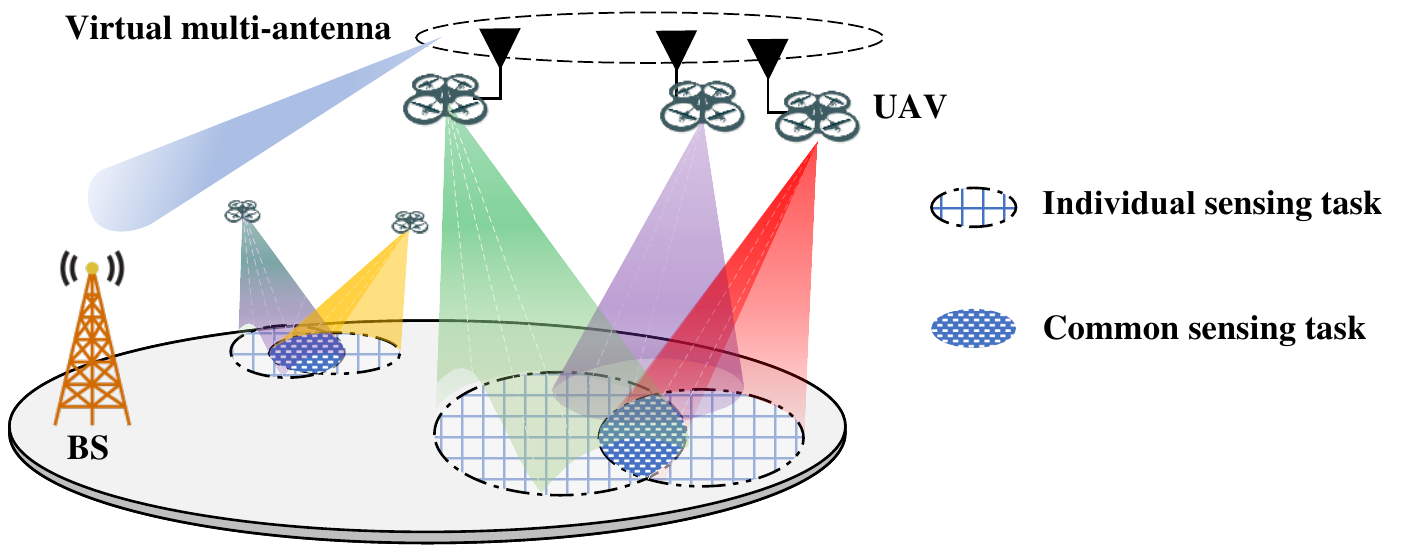}
	\caption{The scenario of multi-UAV performing sensing tasks.}
	\label{figure1}
\end{figure}

Motivated by the above, a multi-UAV cooperative S\&T mechanism with overlapped task allocation is proposed in this work. Specifically, in the proposed scheme, part of the mission (termed \textit{common task}) is repetitively allocated to all the UAVs. In this way, the UAVs can form a virtual multiple-input and single-output (MISO) system \cite{VirtualMIMOUserGrouping} \textit{without} using extra communications to share sensing data among each other, and cooperatively transmit the sensory data of the common task to the BS, as shown in Fig.~\ref{figure1}. To the best of our knowledge, this work is among the first to reveal the benefits of overlapped sensing task allocation for multi-UAV cooperative S\&T systems.
With proper optimization, the gain in data transmission rate will override the loss due to overlapped task allocation, eventually leading to a reduction of the overall mission completion time. To minimize the mission completion time, the task allocation and transmit power of the proposed scheme are jointly optimized under energy budget constraints. Nevertheless, solving the resulting problem (c.f. (P1) in Section II-B) is highly non-trivial. First, to increase the transmission rate, one may increase the portion of the common task to promote cooperative transmission, which however, may unfavorably increase the total workload (e.g., sensing time) and energy consumption \cite{Zhan2020Aerial}. This implies a fundamental trade-off between sensing time and transmission time, and hence the portion of the common task needs to be optimized properly. Furthermore, task assignment/association and transmit power allocation are tightly coupled and lead to non-convex constraints. In addition, due to the information-causality constraint of sensory data, the objective function has a complicated multi-level nesting structure. To overcome the above technical challenges, a necessary condition for multi-UAV overlapped sensing is analyzed. For the general cases of overlapped sensing, the resulting problem is transformed into a monotonic optimization (MO) and is solved by the generic Polyblock algorithm. To efficiently evaluate the mission completion time in each iteration of the Polyblock algorithm, new auxiliary variables are introduced to decouple the sophisticated joint optimization of transmission time and power. While for the degenerated case of non-overlapped sensing, an efficient double-loop binary search algorithm is proposed to optimally solve the resulting problem and the closed-form expression of the optimal transmission time is derived.
\par
The main contributions of this work are summarized as follows:
\begin{itemize}
	\item We propose a novel multi-UAV cooperative S\&T scheme to minimize the mission completion time. This scheme reveals the potential benefits of overlapped sensing for multi-UAV cooperative transmission, which implies an important insight into sensing-assisted communication and provides more flexibility for balancing sensing and communication in multi-UAV cooperative S\&T schemes. 
	\item To solve this problem, we derive a necessary condition for overlapped sensing. For the general overlapped sensing cases, the problem is transformed into an MO problem, where the transmit power is optimally obtained by introducing new auxiliary variables and decoupling the transmit time and power. For the degenerated case of non-overlapped sensing, the optimal transmission time for a given task allocation is derived in closed-form, and a double-loop binary search algorithm is proposed to optimally solve the degenerated problem.   
	\item Finally, through simulations, it is observed that the UAVs tend to raise the portion of the common task when the sensing data acquisition time is short and the energy budget is high, and also that the proposed scheme can significantly reduce the overall mission completion time. 
\end{itemize}
\par
The rest of this paper is organized as follows. In Section \ref{SYSTEM}, the problem formulation and system model are presented. In Section \ref{SufficientCondition}, a necessary condition of overlapped sensing and a sufficient condition of fully overlapped sensing are analyzed. The proposed algorithms for the cooperative S\&T scheme are developed in Section \ref{OptimalSolution}. In Section \ref{simulation}, simulation results are presented. Finally, conclusions and future works are discussed in Section \ref{Conclusion}.

\section{Problem Formulation and System Model}
\label{SYSTEM}
In this section, the proposed multi-UAV cooperative S\&T scheme is presented first, followed by the formulation of the mission completion time minimization problem. Important notations and symbols used in this work are given in Table \ref{Notation}.
\begin{table}[h]
	\small
	\caption{Important notations and symbols.} 
	\label{Notation}	
	\centering
	\begin{tabular}{ll}%
		\hline
		$\rm{\textbf{Notation}}$  & $\rm{\textbf{ Physical meaning}}$ \\
		\hline
		$T$                       &  Total mission completion time  \\
		\hline  
		$p_m^c$                   &  Transmit power of cooperative transmission  \\ 
		\hline  
		$p_m^n$                   &  Transmit power of independent transmission \\
		\hline  
		$h_m$                     &  Channel power gain of UAV $m$ \\
		\hline  
		$T^c$                     &  Cooperative transmission time \\
		\hline  
		$T^n_m$                   &  Independent transmission time of UAV $m$\\
		\hline  
		$T^s_m$                   &  Sensing time of UAV $m$\\
		\hline  
		$\beta^s$                 &  Total workload of the entire mission \\
		\hline  
		$\omega_m$                &  Individual sensing task ratio of UAV $m$ \\
		\hline  	
		$\omega_0$                &  Common sensing task ratio \\
		\hline  
		$E_m$                     &  Energy consumption of UAV $m$\\
		\hline  
		$\bar {E}$                &  Transmit energy budget \\
		\hline  	
	\end{tabular}
\end{table}
\subsection{System Model}
A multi-UAV cooperative S\&T system is investigated, as depicted in Fig.~\ref{figure1}, where $M$ UAVs, indexed by $m \in {\cal{M}} = \{1,...,M\}$, cooperatively perform a sensing mission. To reduce the mission completion time, the entire mission is divided into two parts: A common task and $M$ individual sensing tasks. The common task is repeatedly executed by all UAVs, i.e., overlapped sensing task allocation,\footnote{In this work, it is assumed that the UAVs share a common field of view. This is reasonable for many practical scenarios, especially when the UAVs fly at high altitudes and are relatively close to each other. When the sensing area is very large, the UAVs could be further divided into several groups according to whether they can have overlapped fields of view. For each group, the proposed collaborative scheme can be adopted in the same way as presented in this work.} and the corresponding sensory data is transmitted to the BS through the virtual MISO technique \cite{VirtualMIMODataGathering, ZeroForcingVirtualMIMO}.\footnote{In our considered model, multiple UAVs each equipped with a single antenna form the multi-input part of the virtual MISO, and the BS with a single antenna forms the single-output part.} The common task accounts for the $\omega_0$ portion of the entire mission.{\footnote{In this work, it is assumed that the mission can be divided continuously for ease of analysis \cite{Huang2022OPAT, Wang2021SubChannel}.}} The remaining $(1-\omega_0)$ portion of the entire mission is further divided into $M$ individual sensing tasks and allocated to $M$ UAVs. The $m$th individual task constitutes $\omega _m$ portion of the entire mission and will be transmitted by UAV $m$ to the BS independently. \footnote{The original sensing mission can be divided into multiple tasks with proper granularity and these sensing tasks can be indexed based on a certain rule, such as locations, directions, etc. After obtaining the task allocation result from the proposed algorithm, the indices of the common tasks and those of the individual tasks can be sent to the UAVs through a control link. Note that the cost of sending this information could be negligible, since the mapping between the task indices and the actual task locations can be made available to the UAVs beforehand. }

\begin{figure}[t]
	\centering
	\setlength{\abovecaptionskip}{0.cm}
	\includegraphics[width=15cm]{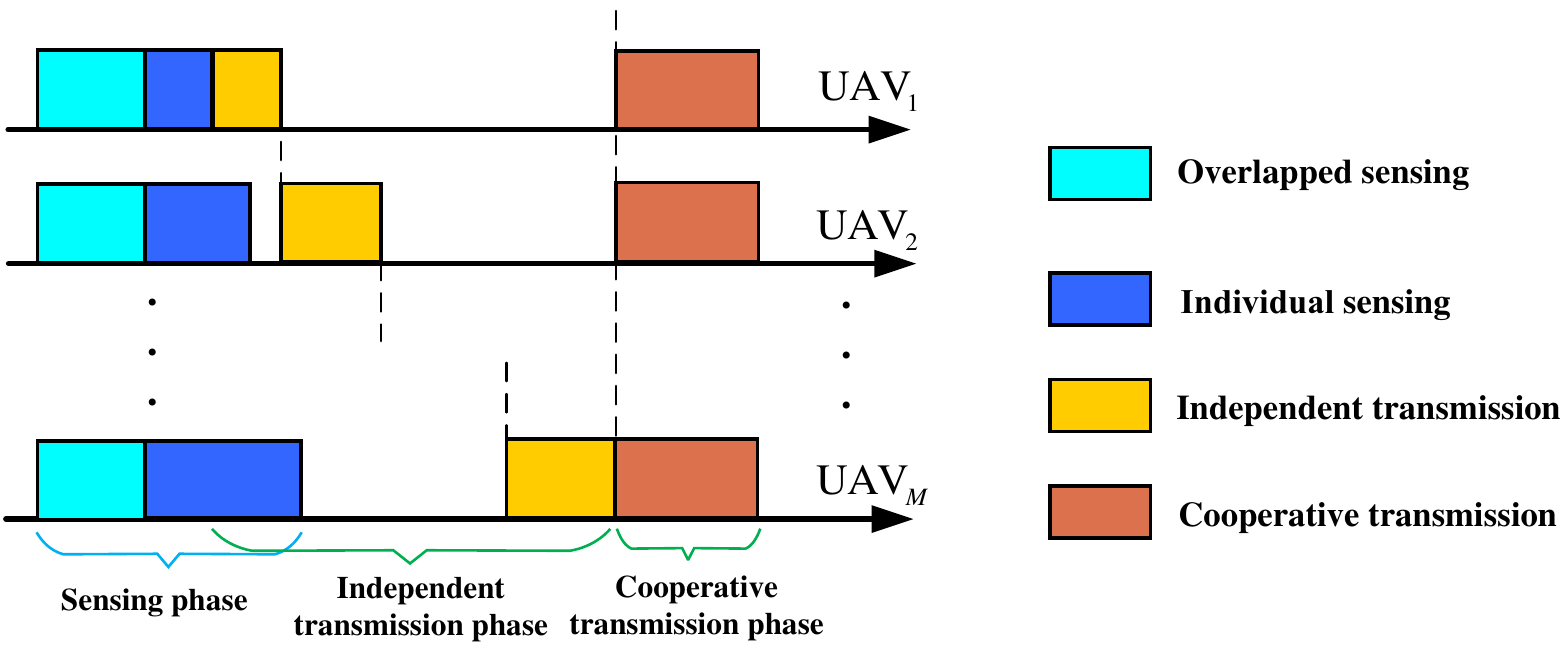}
	\caption{Illustration of the proposed cooperative S\&T scheme with overlapped sensing.}
	\label{figure2}
\end{figure}

The proposed multi-UAV S\&T scheme consists of a sensing phase and two transmission phases, as shown in Fig. 2. It is assumed that all UAVs start sensing simultaneously and the common task will be performed first. The independent transmission phase aims to deliver the data of the individual sensing tasks and begins right after the first UAV completes sensing. To avoid mutual inference among the UAVs, time division multiple access (TDMA) is adopted in the proposed scheme, in which each UAV will start to transmit when its sensing tasks are completed and the channel is idle.{\footnote{In case of competing, the channel will be assigned to the UAV that finishes sensing earlier.}} Note that as shown in Fig.~\ref{figure2}, the sensing phase and the independent transmission phase may overlap. After delivering all the data of the individual sensing tasks, the UAVs begin to transmit the data of the common task cooperatively in the second transmission phase, by forming a virtual MISO. Generally, to reduce the mission completion time, the UAV with larger channel power gain tends to perform more tasks. Hence, it is assumed that the UAVs are indexed based on their channel power gains in our proposed scheme and the task allocation satisfies $\omega_{m} \le \omega_{m+1}$, $\forall m \in {\cal{M}} \backslash \{M\}$. With the aforementioned model, it is not difficult to verify that, the total mission completion time $T$ of the proposed scheme is given by
\begin{equation}\label{MissionCompletionTime}
T = \underbrace {\max \left( {\max \left( {...,T_{M - 1}^s} \right) + T_{M - 1}^n,T_M^s} \right)}_{
	{M - 1} \quad \text{nested} \quad {\max \left( \cdot , \cdot \right)}
	} + T_M^n + {T^c},
\end{equation}
where $T_m^{s}$ is the total {\textit{sensing time}} of UAV $m$ including both the individual tasks and the common tasks, $T_m^{n}$ is the {\textit{independent transmission time}} of UAV $m$ for delivering the data of its individual tasks, and $T^{c}$ is the {\textit{cooperative transmission time}} of all UAVs for delivering the data of the common task.
\par
In this work, it is assumed that the sensing time is generally proportional to the amount of tasks \cite{Wang2018MultiTask}, and hence, the total sensing time of UAV $m$ can be modeled as
\begin{equation}
T_m^s = \left( {{\omega _m} + {\omega _0}} \right) {\beta^s},
\end{equation} 
where $\beta^s$ represents the total required time (measured in seconds) of performing the entire mission by a single UAV, namely workload. The power gain of the wireless channel from UAV $m$ to the BS is denoted by $|{h_{m}}|^2$ \cite{CellularEnabledUAVCommunication}. When UAV $m$ transmits independently, the corresponding SNR is given by ${{{p^n_m} |{h_{m}}|^2} \mathord{\left/{\vphantom {{{p^n_m} |{h_{m}}|} {{\sigma ^2}}}} \right.\kern-\nulldelimiterspace} {{\sigma ^2}}}$, where ${p^n_m}$ is the transmit power of UAV $m$ for its individual tasks and ${\sigma ^2}$ is the variance of the channel noise.\footnote{In this work, we mainly consider the sensors that do not interfere with communications, such as cameras, Lidar, or radar sensors operating in separate frequency bands from communications.} In the subsequent discussions, let ${\gamma _m} = {{ |{h_{m}}|^2} \mathord{\left/{\vphantom {{ |{h_{m}}|} {{\sigma ^2}}}} \right.\kern-\nulldelimiterspace} {{\sigma ^2}}}$ for notation convenience. Hence, the independent transmission rate from UAV $m$ to the BS can be expressed as \cite{Wu2018JointTrajectory}
\begin{equation}\label{Rate_m2BS}
r^n_{m} = B {{\log }_2}\left( {1 + {p^n_m} {\gamma}_{m}} \right),
\end{equation}
where $B$ is the bandwidth. Accordingly, the time of transmitting 1-bit of sensory data from UAV $m$ to the BS is given by 
\begin{equation}\label{TransmissionTimeNonCooperative}
t^n_m = \frac{1}{{B  {{\log }_2}\left( {1 + {p^n_m} {\gamma _{m}}} \right)}}.
\end{equation}
In this work, it is assumed that the amount of data is proportional to the number of tasks. Hence, it takes $T_m^n = {\omega _m}  C  {t_m^n}$ seconds for UAV $m$ to transmit the data of its individual sensing tasks, where $C$ represents the amount of sensory data of the entire mission. On the other hand, in the cooperative transmission phase, the data rate from $M$ UAVs to the BS via virtual MISO is given by \cite{VirtualMIMO5G, DownlinkMIMO, tse2005fundamentals}
\begin{equation}
r^c = B  {{{\log }_2} \left(1 +  {{\sum\nolimits_{m = 1}^M{p_m^c  {\gamma_{m}}}}} \right)},
\end{equation}
where $p_m^c$ is the transmit power of UAV $m$ for cooperative transmission. In this work, it is assumed that the UAVs and the BS are synchronized, as existing synchronization methods for conventional wireless communications \cite{li2008timing, tian2017time} may be applied. In this case, interaction among the UAVs is not required, since all UAVs can start to perform the sensing tasks and transmit the sensory data based on the timeline specified by the proposed scheme. The corresponding transmission time of 1-bit sensory data is given by 
\begin{equation}\label{TransmissionTimeCooperative}
{t^c}{\rm{ = }}\frac{1}{{B  {{\log }_2}\left( {1 + {{\sum\nolimits_{m = 1}^M {p_m^c  {\gamma_{m}}} }}} \right)}}.
\end{equation}
Hence, the cooperative transmission time is $T^c = \omega_0  C  {t^c}$. The total energy consumption of UAV $m$ in these two transmission phases is given by
\begin{equation}
E_m = \omega_0  C  t^c  {p^c_m} + {\omega _m}  C  t^n_m  {p^n_m}.
\end{equation}

\subsection{Problem Formulation}
In this work, we aim to minimize the mission completion time by optimizing task allocation and transmit power. Mathematically, this problem can be formulated as follows:
\begin{subequations}\label{P1}
\begin{align}
(\rm{P1}): \quad & \begin{array}{*{20}{c}}
\mathop {\min }\limits_{ {\bm{\omega}} ,{\bm{p}}^c ,{\bm{p}}^n } \mathop T 
\end{array}  \\ 
\mbox{s.t.}\quad
\label{P1-a}& 0 \le E_m \le {{\bar {E}}},\forall m \in {\cal{M}}, \\
\label{P1-b}&\sum\nolimits_{m = 0}^M {{\omega _m}}  = 1, \\
\label{P1-c}& 0 \le {\omega _m} \le 1,\forall m \in {\cal{M}} \cup \{0\}, \\
\label{P1-d}& 0 \le p_m^c \le p^{\max },\forall m \in {\cal{M}}, \\
\label{P1-e}& 0 \le p_m^n \le p^{\max },\forall m \in {\cal{M}}.
\end{align} 
\end{subequations}
In (P1), vector ${\bm{\omega}} = \{\omega_m\}_{m=1}^M$ represents the task allocation ratios, vectors ${\bm{p}}^n = \{p_m^n\}_{m=1}^M$ and ${\bm{p}}^c = \{p_m^c\}_{m=1}^M$ denote the transmit power of the UAVs for the individual tasks and the common task, respectively, and $p^{\max}$ is the maximum transmit power of the UAVs. Constraints in (\ref{P1-a}) denote that the total transmit energy per UAV should not exceed $\bar{E}$ \cite{UAVAidedMIMO, CompletionTimeMinimization}. Constraint (\ref{P1-b}) guarantees the completion of the entire mission. Solving (P1) optimally is highly non-trivial due to the non-convex constraints (\ref{P1-a}), the coupled optimization variables, and the nested structure of the non-convex objective function (c.f. (\ref{MissionCompletionTime})). To tackle this issue, a necessary condition for overlapped sensing will be analyzed first, and based on which, (P1) can be solved efficiently as presented in the next section.

\section{Conditions for Overlapped Sensing}
\label{SufficientCondition}

In this section, the necessary conditions for overlapped sensing (i.e., $\omega_0 > 0$) and the sufficient conditions for the special case of fully overlapped sensing (i.e., $\omega_0 = 1$) are analyzed to facilitate subsequent derivations of the optimal task and power allocations. To this end, the following lemma is useful to work around relevant technical difficulties caused by the nested structure of the objective function in (P1).

{\setlength{\parindent}{0em}
	\begin{thm}{\label{EqualProblem1}}
		Problem (P1) can be equivalently rewritten as:
		\begin{subequations}\label{P1.1}
		\begin{align}	
		(\rm{P1.1}): \quad & \mathop {\min }\limits_{{\bm{\omega}} ,{\bm{p}}^n ,{\bm{p}}^c } T_1^s + \sum\nolimits_{m = 1}^M {T_m^n}  + {T^c}   \\ 
		\mbox{s.t.}\quad
		& ({\rm\ref{P1-a}})-(\rm{\ref{P1-e}}),  \nonumber \\
		\label{P1.1-a}& T_m^{\max} + T_m^n \ge T_{m + 1}^s,\forall m \in  {\cal{M}} \backslash \{M\},  
		\end{align} 
		\end{subequations}
		where $T_1^{\max} = T_1^s$, $T_m^{\max} = T_{m-1}^{\max} + T_{m-1}^n$ for $m \in {\cal{M}} \backslash \{1\}$, and $T_m^{\max}$ represents the time instant when UAV $m$ can start data transmission. 
	\end{thm}
}
\begin{proof}
	Please refer to Appendix A.
\end{proof} 

By Lemma \ref{EqualProblem1}, the nested structure in the objective function of (P1) is removed by introducing the auxiliary variables $T_m^{\max}$ and adding constraint (\ref{P1.1-a}). However, it is still challenging to solve (P1.1), since the task allocation ${\bm{\omega}}$, independent transmit power ${\bm{p}}^n$, and cooperative transmit power ${\bm{p}}^c$ in (P1.1) are coupled in the non-convex constraints ({\rm\ref{P1-a}}). To overcome this difficulty, the relation between the optimal independent transmit power $p_m^n$ and the optimal cooperative transmit power $p_m^c$ is derived below.
\par
{\setlength{\parindent}{0em}
	\begin{pro}\label{SufficientConditions}
		At the optimal solution of (P1), if the common task ratio $\omega^*_0 > 0$, the following condition holds:
		\begin{equation}\label{OptimalNonCooeprativepower}
			p_m^n = \left\{ {\begin{array}{*{20}{c}}
					{\min\left(\frac{\sum\nolimits_{m = 1}^M {p_m^c{\gamma _m}}}{\gamma_m}, p^{\max}\right),}&{m = M}\\
					{\min\left( \frac{\sum\nolimits_{m = 1}^M {p_m^c{\gamma _m}}}{\gamma_m}, \min ( {\bar{p}_m^n,{p^{\max }}} )\right) ,}&{{\rm{otherwise}}}
			\end{array}}, \right.
		\end{equation}
	where $\bar {p}_m^n = \frac{1}{{\gamma _m}} \left({{2^{\frac{{C  {\omega _m}}}{{ B {} ({T^s_{m+1} - T_m^{\max}})}}}} - 1}\right), m \in  {\cal{M}} \backslash \{M\}$.
	\end{pro}
}
\begin{proof}
	Please refer to Appendix B.
\end{proof}
\par
By Proposition \ref{SufficientConditions}, the relationship between the transmission times of the common task and the individual tasks can be derived. 

{\setlength{\parindent}{0em}
	\begin{cor}\label{OptimalCooperativeTime}
		At the optimal solution of (P1), if the common task ratio $\omega^*_0 > 0$, the optimal transmission time satisfies ${t^{c*}} \le {t^{n*}_m}$, $\forall m \in {\cal{M}}$.
	\end{cor}
}
\begin{proof}
	Please refer to Appendix C. 
\end{proof} 

Corollary \ref{OptimalCooperativeTime} indicates that if the optimal common task ratio $\omega_0^* > 0$, the cooperative transmission time of 1-bit sensory data is always less than that of the independent transmission time. Based on Corollary \ref{OptimalCooperativeTime}, a necessary condition for overlapped sensing is further derived for problem analysis as follows.
{\setlength{\parindent}{0em}
	\begin{pro}\label{SufficientConditions2}
		The necessary condition for overlapped sensing (i.e., $\omega_0 > 0$) is $x > p^{\max} \gamma_M$, where x can be obtained via binary search for the following equation.
		\begin{equation}\label{CooperativeEquation}
			\frac{{  C  x }}{{B{{\log }_2}(1 + x )}} = \bar {E} \sum\nolimits_{m = 1}^M   \gamma_m.
		\end{equation}
	\end{pro}
}
\begin{proof}
     Please refer to Appendix D.
\end{proof} 
{\setlength{\parindent}{0em}
\begin{remark}
	By Proposition 2, if the solution $x$ of (\ref{CooperativeEquation}) satisfies $ x \leq p^{\max}{\gamma_M}$, it is not necessary to conduct overlapped sensing. In this case, (P1.1) reduces to an optimization problem with only ${\bm{\omega}}$ and ${\bm{p}}^n$. Moreover, Proposition 2 indicates that, when the energy budget $\bar{E}$ is sufficiently large and the amount of sensing tasks $C$ is relatively small, the UAVs tend to cooperate, so as to reduce the mission completion time. Otherwise, they tend to perform tasks independently.
\end{remark}}

Although overlapped sensing enables faster cooperative transmission of sensory data, it increases the overall sensing workload of the UAVs and thus the corresponding sensing time. This implies a fundamental trade-off between the sensing time and the transmission time. Therefore, the workload $\beta^s$ is an important factor in determining the necessity of overlapped sensing in the proposed multi-UAV S\&T scheme. In particular, the following result holds.

{\setlength{\parindent}{0em}
	\begin{pro}{\label{TsEqual0}}
		If $\beta^s \to 0$, the optimal cooperative task ratio $\omega^*_0 = 1$; if $\beta^s \to \infty$, the optimal cooperative task ratio $\omega^*_0 = 0$.
	\end{pro}
}
\begin{proof}
	Please refer to Appendix E. 
\end{proof} 

Proposition \ref{TsEqual0} is consistent to the intuition. If the workload $\beta^s$ is negligible, the overlapped sensing is more time-efficient due to cooperative transmission, and thus fully overlapped sensing, i.e., $\omega^*_0 = 1$, is optimal. On the other extreme that $\beta^s \to \infty$, individual sensing ratio $(\omega_0 = 0)$ is preferred to improve the sensing efficiency.
\par

\section{{Optimal Sensing and Transmission}}
\label{OptimalSolution}
The algorithms for solving (P1) are developed in this section, by exploiting the analytical results in Section \ref{SufficientCondition}. In particular, Proposition {\ref{SufficientConditions2}} will be used to check if overlapped sensing is needed. For the general cases where the necessary condition (\ref{CooperativeEquation}) holds, an MO-based algorithm is proposed. While for the degenerated case where (\ref{CooperativeEquation}) does not hold, it follows that $\omega_0 = 0$ and an efficient double-loop search algorithm is proposed.

\subsection{General Cases}
\label{CooperativeMode}
Some preliminaries about MO are presented first to streamline the subsequent discussions \cite{tuy2000monotonic, zhang2013monotonic}.
\newtheorem{myDef}{\bf Definition}{
	\setlength{\parindent}{0em}
	\begin{myDef}
		(\textit{Conormal}): A set ${\cal{H}} \subset \mathbb{R}_ + ^{N}$ is conormal if ${\bm{x}} \in {\cal{H}}$ and ${\bm{x'}} \succeq {\bm{x}}$ implies ${\bm{x'}} \in {\cal{H}}$. Here, ${\bm{x'}} \succeq {\bm{x}}$ means that ${{\bm{x'}}}$ is component-wise no less than ${{\bm{x}}}$.
	\end{myDef}
	\begin{myDef}
		(\textit{Monotonic optimization}): An optimization problem can be transformed to an MO if it can be written as
		\begin{shrinkeq}{-2ex}
			\begin{alignat}{2}
				\label{MO_define}
				\mathop {{\rm{max}}}\limits_{\bm{x}}  \quad & g(\bm{x}), \nonumber & \\
				\mbox{s.t.}\quad
				&{{\bm{x}} \in \cal{H}}, &  
			\end{alignat} 
		\end{shrinkeq}
		where ${\cal{H}} \subset \mathbb{R}_ + ^{N}$ is a non-empty conormal set and $g(\bm{x})$ is decreasing over $\mathbb{R}_ + ^{N}$.
	\end{myDef}
	\begin{myDef}
		(\textit{Box}): Let $\bm{y},\bm{z} \in \mathbb{R}_ + ^{N}$ and ${\bm{z}} \succeq {\bm{y}}$. The hyper-rectangle determined by $\bm{y}$ and $\bm{z}$ is called a box in $\mathbb{R}_ + ^{N}$, i.e., $\left[ {{\bm{y}},{\bm{z}}} \right] = \prod\limits_{n = 1}^N {\left[ {{y_n},{z_n}} \right]} $, where ${\left[ {{y_n},{z_n}} \right]}$ is the $n$th dimension of the box. The points $\bm{y}$ and $\bm{z}$ are called the lower bound vertex and the upper bound vertex, respectively.
	\end{myDef}
	\begin{myDef}
		(\textit{Copolyblock}): A set ${\cal{Q}} \subset \mathbb{R}_ + ^{N}$ is called a copolyblock if it is a union of a finite number of boxes $[{\bm{y}},{\bm{b}}]$, where ${\bm{y}} \in {\cal{V}}$, i.e., ${\cal{Q}} = { \cup _{{\bm{y}} \in \cal{V}}}\left[ {{\bm{y}},{\bm{b}}} \right]$. Clearly, Polyblock is a normal set.
	\end{myDef}
	\begin{myDef}
		(\textit{Projection}): Given any non-empty normal set ${\cal{H}} \in \mathbb{R}_ + ^{N}$ and any box $\left[ {{\bm{y}},{\bm{z}}} \right]$, $\Phi ({\bm{y}})$ is the projection of $\bm{z}$ onto the boundary of ${\cal{H}}$, $\Phi ({\bm{y}}) = {\bm{y}} + \lambda {(\bm{z}-\bm{y})}$, where $\lambda  = \max \{ \mu |({\bm{y}} + \mu {(\bm{z}-\bm{y})}) \in \cal{H}\} $ and $\mu  \in {\mathbb{R}_ + }$.
	\end{myDef}
}
{\setlength{\parindent}{0em}
\begin{pro}\label{MOFact}
	Problem (P1.1) is an MO (c.f. {\bf{Definition 2}}).
\end{pro}}
\begin{proof}
	The objective function of (P1.1) increases monotonically in ${\bm{\omega}}$, since both the sensing time $\{T^s_m\}_{m=1}^M$ and the transmission time $\{T^n_m\}_{m=1}^M$, increase monotonically with ${\bm{\omega}}$.
\end{proof}
Based on Proposition \ref{MOFact}, (P1.1) can be solved using the Polyblock algorithm \cite{zhang2013monotonic}. Specifically, in the $r$th iteration of the Polyblock algorithm, the vector ${\bm{\omega}}^{(r)} = [\omega_0^{(r)}, \cdots, \omega_M^{(r)}]$ corresponding to the minimum mission completion time is selected, and then its projection point $\Phi\left({\bm{\omega}}^{(r)}\right) = [\Phi_1\left({\bm{\omega}}^{(r)}\right),\cdots,\Phi_{M+1}\left({\bm{\omega}}^{(r)}\right)]$ can be calculated. The $m$th coordinate $\Phi_m\left({\bm{\omega}}^{(r)}\right)$ of the projection point is given by
\begin{equation}\label{ProjectionPoint}
	\Phi_m\left({\bm{\omega}}^{(r)}\right) = \left(\bar \omega_m - \omega_m^{(r)}\right)\frac{1-{\sum\nolimits_{m = 1}^M} \bar \omega_m}{\bar \omega_m - \omega_m^{(r)}} + \omega_m^{(r)},
\end{equation}
where $\bar \omega_m$ denotes the upper bound of the individual task ratio of UAV $m$. After each iteration, a smaller copolyblock ${\cal{Q}}^{(r+1)}$ can be constructed by replacing the vertices ${\bm{\omega}}^{(r)}$ with the newly generated vectors in copolyblock ${\cal{Q}}^{(r)}$. Note that, to find the mission completion time corresponding to each possible allocation vector ${\bm{\omega}}$ in the iterations of the Polyblock algorithm, one needs to solve the following sub-problem:
\begin{subequations}\label{P2}
	\begin{align}	
		(\rm{P2}): \quad & \mathop {\min }\limits_{{\bm{t}}^n, {\bm{p}}^c, t^c } T_1^s + C \sum\nolimits_{m = 1}^M \omega_m t^n_m  + C \omega_0 t^c \\ 
		\mbox{s.t.}\quad
		& ({\rm{\ref{P1-d}}}), \nonumber \\
		\label{P2-a}& C \omega_m  t^n_m  \frac{(2^{\frac{1}{Bt_m^n}} - 1)}{\gamma_m} +   {C \omega_0 p_m^c}t^c  \le \bar E,\forall m \in {\cal{M}}, \\
		\label{P2-b}& \omega_1 \beta^s + C \sum\nolimits_{i = 1}^m \omega_i t_i^n \ge T_{m + 1}^s,\forall m \in {\cal{M}} \backslash \{M\}, \\
		\label{P2-c}& 0 \le t_m^n \le t_{m,\min}^n, \forall m \in {\cal{M}}, \\
		\label{P2-d}& 0 \le t^c \le t_{\min}^c,
	\end{align}	
\end{subequations}
where ${\bm{t}}^n = \{t_m^n\}_{m=1}^M$ is a compact representation of the independent transmission times of 1-bit sensory data, $t_{m,\min}^n = \frac{1}{B \log_2(1+ p^{\max} \gamma_m)}$, and $t_{\min}^c = \frac{1}{B \log_2(1+ \sum\nolimits_{m = 1}^M p^{\max} \gamma_m)}$. To resolve the non-convex constraints (\ref{P2-a}) and (\ref{P2-b}) in (P2), new auxiliary variables $\{E^c_m\}$, which represents the energy budget of cooperative transmission, are introduced to decouple the cooperative transmit power $p_m^c$ and transmission time $t^c$. As such, (P2) can be equivalently written as
\begin{subequations}\label{P2.1}
	\begin{align}
		(\rm{P2.1}): \quad & \mathop {\min }\limits_{{\bm{t}}^n, {\bm{p}}^c, t^c, \{E_m^c\} } T_1^s + C \sum\nolimits_{m = 1}^M \omega_m t^n_m  + C \omega_0 t^c \\ 
		\mbox{s.t.}\quad
		& ({\rm{\ref{P1-d}}}), ({\rm{\ref{P2-b}}})-({\rm{\ref{P2-d}}}), \nonumber \\
		\label{P2.1-a} & C \omega_m  t^n_m  \frac{(2^{\frac{1}{Bt_m^n}} - 1)}{\gamma_m} +   E_m^c  \le \bar E,\forall m \in {\cal{M}}, \\
		\label{P2.1-b} & C \omega_0 p_m^c t^c \le E_m^c, \forall m \in {\cal{M}},  \\
		\label{P2.1-c} & 0 \le E_m^c \le \bar {E}.
	\end{align}	
\end{subequations}
It can be readily proved that $ t^n_m (2^{\frac{1}{Bt_m^n}} - 1)$ is convex about $t^n_m$ for $t^n_m \ge 0$. Although the constraints in ({\ref{P2.1-b}}) are still non-convex, they can be transformed into a form only related to $t^c$ according to the definition in (\ref{TransmissionTimeCooperative}). Specifically, multiplying both sides of (\ref{P1-a}) with $\gamma_ m$ and accumulating these energy consumption constraints yield
\begin{equation}
	C \omega_0  t^c   (2^{\frac{1}{Bt^c}} - 1) \le \sum\nolimits_{m = 1}^M  E_m^c \gamma_m.
\end{equation}
Furthermore, plugging (\ref{TransmissionTimeCooperative}) into (\ref{P2.1-b}) leads to the following equivalent form of (P2.1)
\begin{alignat}{2}
	\label{P2.2}
	(\rm{P2.2}): \quad & \mathop {\min }\limits_{{\bm{t}}^n, t^c, \{E_m^c\} } T_1^s + C \sum\nolimits_{m = 1}^M \omega_m t^n_m  + C \omega_0 t^c & \\ 
	\mbox{s.t.}\quad
	& ({\rm{\ref{P2-b}}})-({\rm{\ref{P2-d}}}), ({\rm{\ref{P2.1-a}}}), ({\rm{{\ref{P2.1-c}}}}),  \nonumber \\
	& C \omega_0  t^c   (2^{\frac{1}{Bt^c}} - 1) \le \sum\nolimits_{m = 1}^M  E_m^c \gamma_m, & \tag{\ref{P2.2}a}  \\
	& \omega_0 C t^c p^{\max} \ge E_m^c, \forall m \in {\cal{M}}, & \tag{\ref{P2.2}b} 
\end{alignat} 
where constraint ({\ref{P2.2}b}) ensures that a unique optimal transmit power $p_m^c$ can be recovered based on the obtained optimal solution of (P2.2), i.e., $p_m^c = \frac{E_m^c}{C \omega_0 t^c}$, which is also feasible for (P2.1) (and hence (P2)). On the other hand, it can be readily proved that the optimal solution of (P2.1) satisfies the constraints in (P2.2). Therefore, the minimum mission completion time under any given task allocation can be obtained by only finding the optimal transmission times $\{t_m^n\}$ and $t^c$, thus reducing the algorithm complexity.

With the above discussions, (P1.1) can be solved via {\bf{Algorithm} \ref{MONObasedAlgorithm}} as follows. In the outer loop, a feasible task allocation vertex $\{\omega_m\}_{m=0}^M$ is obtained via point projection (c.f. {\bf{Definition 5}}) while generating a smaller copolyblock. For each $\{\omega_m\}_{m=0}^M$ in the outer loop, the corresponding optimal transmission times $\{t_m^n\}$ and $t^c$ are obtained by solving (P2.2) via the standard convex optimization solvers \cite{Michael2014cvx}. 
\begin{algorithm}[t]
	\caption{Monotonic Optimization-Based Algorithm}
	\small
	\label{MONObasedAlgorithm}
	\begin{algorithmic}[1]
		\STATE Set $r = 0$, $\bar \omega_m = \min \left(1,\frac{1}{M+1-m}\right)$ and  $\omega_m^{(r)}=0$, put vector ${\bm{\omega}}$ into vertex set ${{\cal{W}}^{(r)}}$, and construct copolyblock ${\cal{Q}}^{(r)}$ based on ${{\cal{W}}^{(r)}}$. 
		\STATE Let ${\rm{CBV}}^{(r)} = \infty$ and set the convergence accuracy as $\epsilon$.				
		\WHILE {$\frac{|{\rm{CBV}}^{(r)}-\underline{T}^{(r)}|}{{\rm{CBV}}^{(r-1)}} > \epsilon$}
		\STATE Select the vertex ${\bm{\omega}}^{(r)}$ with the minimum mission completion time from vertex set ${\cal{W}}^{(r)}$. Calculate projection point $\Phi({\bm{\omega}}^{(r)})$ according to (\ref{ProjectionPoint}).
		\STATE Obtain ${\rm{CBV}}^{(r+1)} = {T(\Phi({\bm{\omega}}^{(r)}))}$ by solving (P2.2) by CVX, where ${T(\Phi({\bm{\omega}}^{(r)}))}$ is the mission completion time under task allocation $\Phi({\bm{\omega}}^{(r)})$.
		\STATE Construct a smaller copolyblock ${\cal{Q}}^{(r+1)}$ by replacing vertices ${\bm{\omega}}^{(r)}$ in ${{\cal{W}}^{(r)}}$. 	
		\STATE Find vertex 
		${{ {\bm{\omega}}}^{(r+1)}} = \mathop {\arg \min }\limits_{{ {\bm{\omega}}} \in {{\cal{W}}^{(r+1)}}} \left\{ {T({ {\bm{\omega}}})} \right\}$, and the lower bound of task completion is $\underline{T}^{(r+1)} = T({{ {\bm{\omega}}}^{(r+1)}})$. 
		\STATE $r = r+1$.
		\ENDWHILE
		\STATE Recover cooperatively transmit power, i.e., $p_m^c = \frac{E_m^c}{C \omega_0 t^c}$.
	\end{algorithmic}
\end{algorithm}

\subsection{Degenerated Cases}
\label{NonCooperativeMode}
In the degenerated case (i.e., $\omega_0=0$), (P1.1) reduces to
\begin{subequations}\label{P3}
\begin{align}	
(\rm{P3}): \quad & \begin{array}{*{20}{c}}
\mathop {\min }\limits_{{\bm{\omega}}, {\bm{p}}^n } \omega_1 \beta^s  + \sum\nolimits_{m = 1}^M {T_m^n}
\end{array} & \\ 
\mbox{s.t.}\quad
&  ({\rm{\ref{P1-e}}}), (\rm{\ref{P1.1-a}}),  \nonumber \\
\label{P3-b}&\frac{{\omega _m}  C  {p^n_m}}{B \log_2 \left( 1+ p_m^n \gamma_m \right)} \le \bar {E}, \forall m \in {\cal{M}}, \\
\label{P3-c}&\sum\nolimits_{m = 1}^M {{\omega _m}}  = 1, \\
\label{P3-d}& 0 \le {\omega _m}  \le  1	,\forall m \in {\cal{M}},
\end{align}	
\end{subequations}
where with a slight abuse of notation, ${\bm{\omega}} = \{\omega_m\}_{m=1}^M$ is used to represent the task allocation in the degenerated case with after dropping $\omega_0$. To solve this non-convex problem (P3), the following result is useful.

{\setlength{\parindent}{0em}
	\begin{thm}\label{T_function_w}
		Problem (P3) can be equivalently transformed to the following problem:
		\begin{subequations}\label{P3.1}
		\begin{align}
		(\rm{P3.1}): \quad & 
		\mathop {\min }\limits_{{\bm{\omega}} } T_1^s +  \sum\nolimits_{m = 1}^{M} {  C  {\omega _m}  {{\tau}_m}({\omega _m})}  \\ 
		\mbox{s.t.}\quad
		& (\rm{\ref{P3-c}}), (\rm{\ref{P3-d}}),  \nonumber \\
		\label{P3.1-b}& T^s_{m+1} - T^{\max}_m \le { C  {\omega _m} {{\tau}_{m}}({\omega _m})}, \forall m \in {\cal{M}} \backslash \{1\}.
		\end{align}	
		\end{subequations}
		In (P3.1), ${{\tau}_{m}}({\omega _m})$ represents the optimal transmission time of 1-bit sensory data under the energy budget constraints and is given by
		\begin{equation}\label{Expression_T_m}
		{{\tau}_{m}}({\omega _m}) = \left\{ {\begin{array}{*{20}{c}}
			{\frac{{ - \ln 2}}{{B  \left( {W_{-1}\left( { - A_m{\omega _m}  {e^{ - A_m{\omega _m}}}} \right) + A_m{\omega _m}} \right)}},}&{{\omega _m} \ge {{\hat \omega }_m}}\\
			{\frac{1}{{B  {{\log }_2}\left( {1 + {p^{\max }}{\gamma _m}} \right)}},}&{{\omega _m} < {{\hat \omega }_m}}
			\end{array}} \right.,  
		\end{equation}
		where $A_m = \frac{{C \ln 2  }}{{B  {\gamma _m}  {{\bar {E}}}}}$, ${{\hat \omega }_m}{\rm{ = }}\frac{{B{{\log }_2}\left( {1 + {p^{\max }}{\gamma _m}} \right)  {{\bar {E}}}}}{{{p^{\max } C}}}$, and $W_{-1}(\cdot)$ is the Lambert-W function \cite{corless1996lambertw}. 		
	\end{thm}
}
\begin{proof}
	Please refer to Appendix F.
\end{proof}
\par
By Lemma \ref{T_function_w}, (P3) is converted into a single-variable optimization about task allocation ${\bm{\omega}}$. However, the resulting problem (P3.1) is still non-convex due to the non-convex constraint (\ref{P3.1-b}). To tackle (P3.1), an efficient algorithm is proposed by exploiting its inherent monotonic properties.

{\setlength{\parindent}{0em}
	\begin{thm}\label{OptimalCondition}
		At the optimal solution of (P3.1), for UAV $m \in {\cal{M}} \backslash \{M\}$, one of the following conditions holds:
		\begin{itemize}
			\item $T^s_{m+1} - T^{\max}_m = C {\omega^* _m}{{\tau}_{m}}({\omega^* _m}) $ and $\frac{\partial {{\cal{T}}(\omega_m)}}{{{\partial \omega _m}}} |_{\omega_m^*} \ge \frac{{\partial {{\cal{T}}(\omega_m)}}}{{{\partial \omega _{m + 1}}}}|_{\omega_{m+1}^*}$;
			\item $T^s_{m+1} - T^{\max}_m < C { } {\omega^* _m}{{\tau}_{m}}({\omega^* _m}) $ and $\frac{{\partial {\cal{T}}(\omega_m)}}{{{\partial \omega _m}}}|_{\omega_m^*} = \frac{{\partial {{\cal{T}}(\omega_m)}}}{{{\partial \omega _{m + 1}}}}|_{\omega_{m+1}^*}$,
		\end{itemize}
		where $\{\omega_m^*\}$ is the optimal task allocation of UAV $m$, ${{\cal{T}}(\omega_m)} = T_1^s +  \sum\nolimits_{m = 1}^{M} { C {\omega _m}  {{\tau}_{m}}({\omega _m})}$, $\frac{\partial {\cal{T}}(\omega_1)}{{\partial{\omega _1}}} = \beta^s + \frac{{ - {W_{ - 1}}\left( {{x_1}  {e^{{x_1}}}} \right)}}{{\left( {{W_{ - 1}}\left( {{x_1}  {e^{{x_1}}}} \right) - {x_1}} \right)\left( {1 + {W_{ - 1}}\left( {{x_1}  {e^{{x_1}}}} \right)} \right)}} $, $\frac{\partial {\cal{T}}(\omega_m)}{{\partial{\omega _m}}} = \frac{{ - {W_{ - 1}}\left( {{x_m}  {e^{{x_m}}}} \right)}}{{\left( {{W_{ - 1}}\left( {{x_m}  {e^{{x_m}}}} \right) - {x_m}} \right)\left( {1 + {W_{ - 1}}\left( {{x_m}  {e^{{x_m}}}} \right)} \right)}} $ for $m > 1$, and ${x_m} =  - \frac{{C \omega_m \ln 2   }}{{B  {\gamma _m}  {{\bar {E}}}}}$.
	\end{thm}
}

\begin{proof}
	Please refer to Appendix G.
\end{proof}

In Lemma \ref{OptimalCondition}, ${\cal{T}}(\omega_m)$ represents the mission completion time under a given task allocation $\omega_m$. According to Lemma \ref{OptimalCondition}, the optimal task allocation ${\bm{\omega}}$ of (P3.1) could be obtained by comparing the mission completion time ${{\cal{T}}(\omega_m)}$ and its corresponding derivative with respect to (w.r.t) $\omega_m$ of different UAVs. To further facilitate the search of the optimal task allocation, the following result is useful.
{\setlength{\parindent}{0em}
	\begin{pro}\label{MonoIncreasing}
	When $\omega_1$ increases, the task ratios of other UAVs $\{\omega_m\}_{m=2}^M$ admitting Lemma \ref{OptimalCondition} will increase, and the corresponding mission completion time will also increase monotonically with $\omega_1$.
\end{pro}}

\begin{proof}
	Please refer to Appendix H.
\end{proof}

Proposition \ref{MonoIncreasing} implies that, (P3.1) can be solved via a double-loop binary search as in {\bf{Algorithm} \ref{NonCooperativeTransmission}}. Specifically, in the outer loop, a binary search of $\omega_1$ is conducted over the interval $[0,\frac{1}{M}]$. For each $\omega_1$ found in the outer loop, the corresponding optimal mission completion time $T$ can be obtained by searching the optimal task allocation based on Lemma \ref{OptimalCondition} in the inner loop, so as to check whether condition (\ref{P3-c}) in (P3.1) is satisfied. 
\par 
To solve the equation $\frac{{\partial {\cal{T}}(\omega_m)}}{{{\partial \omega _m}}}|_{\omega_m^*} = \frac{{\partial {{\cal{T}}(\omega_m)}}}{{{\partial \omega _{m + 1}}}}|_{\omega_{m+1}^*}$ more efficiently, the relationship among the optimal task ratios of the UAVs is analyzed. 
\par
{\setlength{\parindent}{0em}
	\begin{pro}\label{NonCooperativeOmega_Relation}
		If there exists an optimal solution of (P3.1) in the vector set ${\cal{P}} = \{{\bm{\omega}} |\omega_{m} > {\hat \omega}_{m}, T_{m+1}^s - T_{m}^{\max} \le T_m^n, \forall m\}$, the following conditions hold:
		\begin{equation}\label{Relationship_BetweenW}
			\frac{{{\omega _m}}}{{{\gamma _m}}} = \frac{{{\omega _{m'}}}}{{{\gamma _{m'}}}}.
		\end{equation}
	\end{pro}
}
\begin{proof}
	Please refer to Appendix I.
\end{proof}

{\setlength{\parindent}{0em}
\begin{remark}
	Proposition \ref{NonCooperativeOmega_Relation} provides an intuitive task allocation principle, i.e., more tasks tend to be allocated to UAVs with better channels. The main reason is that with the same energy budget and the amount of sensing tasks, the transmission time of the UAV with a better channel is lower than that with a worse channel.
\end{remark}}

\begin{algorithm}[t]
	\small
	\caption{Double-Loop Binary Search Algorithm}
	\label{NonCooperativeTransmission}
	\begin{algorithmic}[1]
		\STATE Input channel gain $\{\gamma_m\}$, sensing task parameter $C$ and $T_s$.
		\STATE Initialize $r = 1$, ${\bar \omega }_1 = \frac{1}{M}$, and $\underline{\omega}_1 = 0$. 
		
        \WHILE {$| \bar \omega_1   -  \underline{\omega}_1 | \ge \varepsilon $}
		\STATE $\omega_1^{(r)} = \frac{\bar \omega_1  + \underline{\omega}_1}{2}$.
		\FOR {$m = 1:M-1$}
		
		\STATE Calculate $\omega^{(r)}_{m}$ by solving the equation $T^s_{m+1} - T^{\max}_m = \omega_m  {\tau}_{m}(\omega_m)$ for $m \in \{2,...,M-1\}$, or $T^s_2 - T^s_1 = \omega_1  T_1(\omega_1)$.
		\STATE Let $T = \omega_m  {\tau}_{m}(\omega_m)$ for $m \in \{2,...,M-1\}$, and $T = T_m^s + \omega_m  {\tau}_{m}(\omega_m)$ for $m = 1$ according to (\ref{Expression_T_m}).
		\IF {$T_{m+1}^s - T_{m}^{\max} \le T_m^n$}
		\STATE Calculate $\omega^{(r)}_{m}$ based on Proposition \ref{NonCooperativeOmega_Relation}. 
		\ENDIF
		\ENDFOR
		
		\IF {$ {\sum\nolimits_{m = 1}^M {{\omega _m}} > 1}$}		
		\STATE $ {\bar \omega _1} = \omega_1^{(r)}$.		
		\ELSE  		
		\STATE $ {\underline \omega _1} = \omega_1^{(r)}$.		
		\ENDIF
		\STATE $r = r + 1$.
		\ENDWHILE
		
	\end{algorithmic}
\end{algorithm}

\subsection{Computational Complexity Analysis}
In this subsection, the complexity of the proposed S\&T scheme is analyzed. Specifically, the proposed scheme consists of three parts: the necessity check for overlapped sensing, the MO-based algorithm, and the double-loop binary search algorithm. The complexity of verifying the necessity for overlapped sensing is ${\cal{O}}(1)$. The main complexity of the MO-based algorithm comes from solving (P2.2), i.e., step 5. In this step, the complexity of computing the optimal mission completion time is ${\cal{O}}\left((2M + 1)^{3.5}\right)$ \cite{zhang2019securing}, where $2M + 1$ stands for the number of variables. Therefore, the total complexity of the MO-based algorithm is ${\cal{O}}( L_{outer}(M (2M+1)^{3.5}))$, where $L_{outer}$ denotes the number of outer loop iterations required for convergence. The complexity of the double-loop binary search algorithm is ${\cal{O}}(\left( \log _2(n) \right)^2)$. Hence, the complexity of the overall procedure is ${\cal{O}}\left(L_{outer}(M (2M+1)^{3.5}) + \left( \log _2(n) \right)^2\right)$.

\section{{Simulation Results}}
\label{simulation}
In this section, simulation results are provided to evaluate the performance of the proposed scheme. Unless otherwise stated, the simulation parameters are set as follows. The maximum transmit power and the energy budget are set to $p^{\max} = 10$ mW and $\bar {E} = 1$ J, respectively. In addition, the total workload $\beta^s = 2$ s, and the amount of sensory data $C = 20$ Mbits. The bandwidth is set to $B$ = 100 kHz, the number of UAVs $M = 3$, and $\{\gamma _1, \gamma _2, \gamma _3\}$ of these UAVs are set to $\{9 \times 10^3, 1.2 \times 10^4, 1.5 \times 10^4\}$. Besides, the following baselines are considered for comparison:
\begin{itemize}
	\item {\bf{Uniform task allocation without cooperation (UTA-WC)}}: The entire mission is equally divided into $M$ parts, i.e., $\omega_m = \frac{1}{M}$, $\forall m \in {\cal{M}}$, and the common task ratio is set to $\omega_0 = 0$.
	\item {\bf{Uniform task allocation with cooperation (UTA-C)}}: The entire mission is equally divided into $M$ individual tasks and a common task with equal size, i.e., $\omega_m = \frac{1}{M+1}$, $\forall m \in {\cal{M}} \cup \{0\}$.
	\item {\bf{Full cooperation (Full-C)}}: Each UAV performs the entire mission, i.e., $\omega_m = 0$, $\forall m \in {\cal{M}}$, and $\omega_0 = 1$, and all UAVs cooperatively transmit the sensory data.
	\item {\bf{Optimal Allocation without cooperation (Opt-WC)}}: The UAVs take the optimal task allocation and power control without cooperative transmission, i.e., $\omega_0 = 0$.
\end{itemize}
The optimal transmit power in the above benchmarks can be obtained by solving problem (\rm{P2.2}) via CVX tools.
\begin{figure*}[t]
	\centering
	\subfigure
	{	
		\label{figure40}
		\includegraphics[width=14cm]{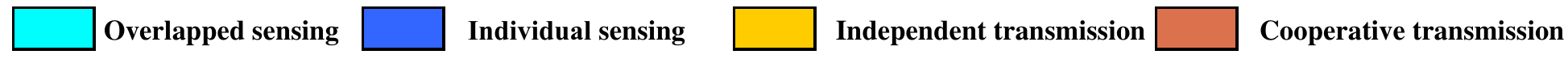}
	}\hspace{6mm}
	\setcounter{subfigure}{0}
	\subfigure[Mission completion time with $\bar {E} = 1$ J and $\beta^s = 4$ s.]
	{	
		\label{figure4a}
		\includegraphics[width=6.9cm]{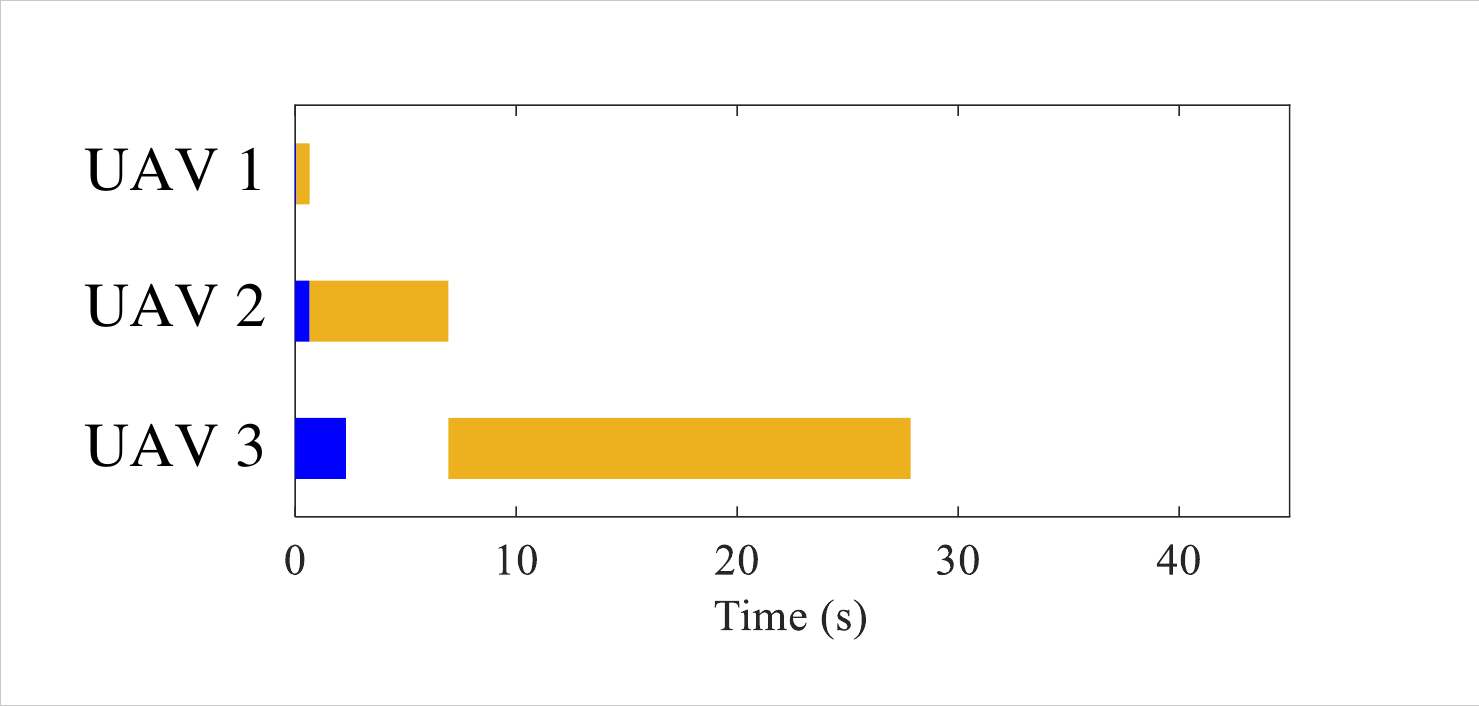}
	}\hspace{6mm}
	\subfigure[Mission completion time with $\bar {E} = 0.1$ J and $\beta^s = 4$ s.]
	{	
		\label{figure4b}
		\includegraphics[width=6.9cm]{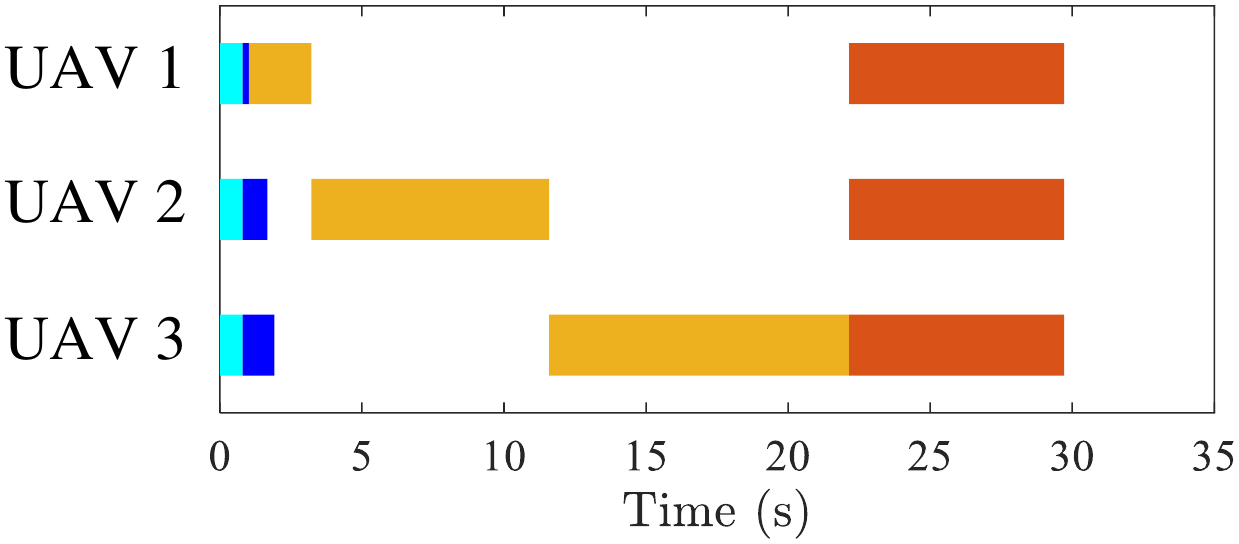}
	}	
	\subfigure[Mission completion time with $\bar {E} = 1$ J and $\beta^s = 10$ s.]
	{	
		\label{figure4c}
		\includegraphics[width=6.9cm]{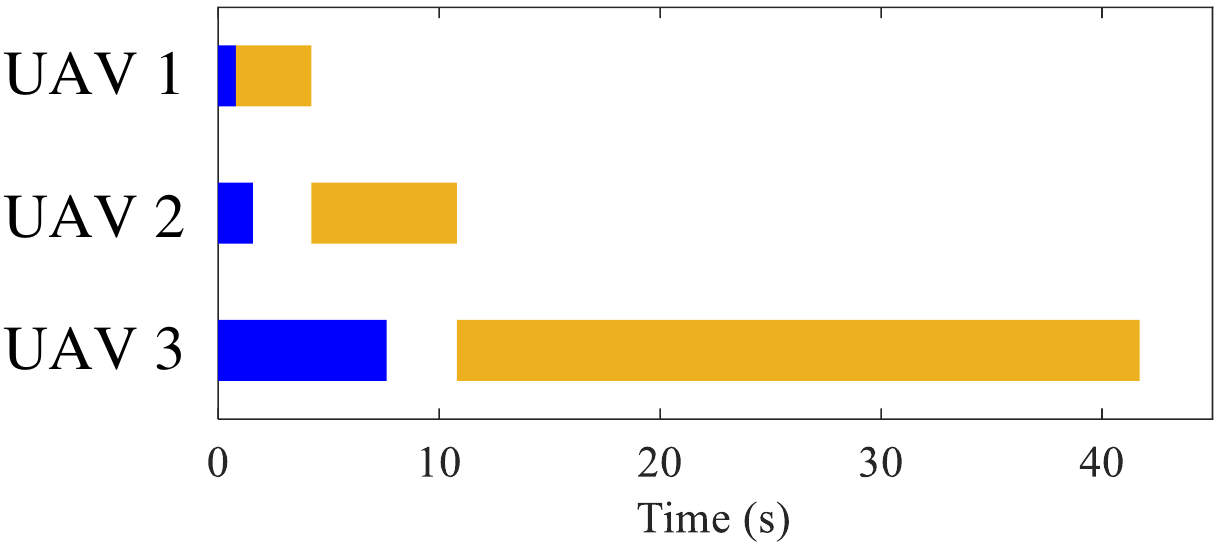}
	}\hspace{6mm}
	\subfigure[Mission completion time with $\bar {E} = 0.1$ J and $\beta^s = 10$ s.]
	{	
		\label{figure4d}
		\includegraphics[width=6.9cm]{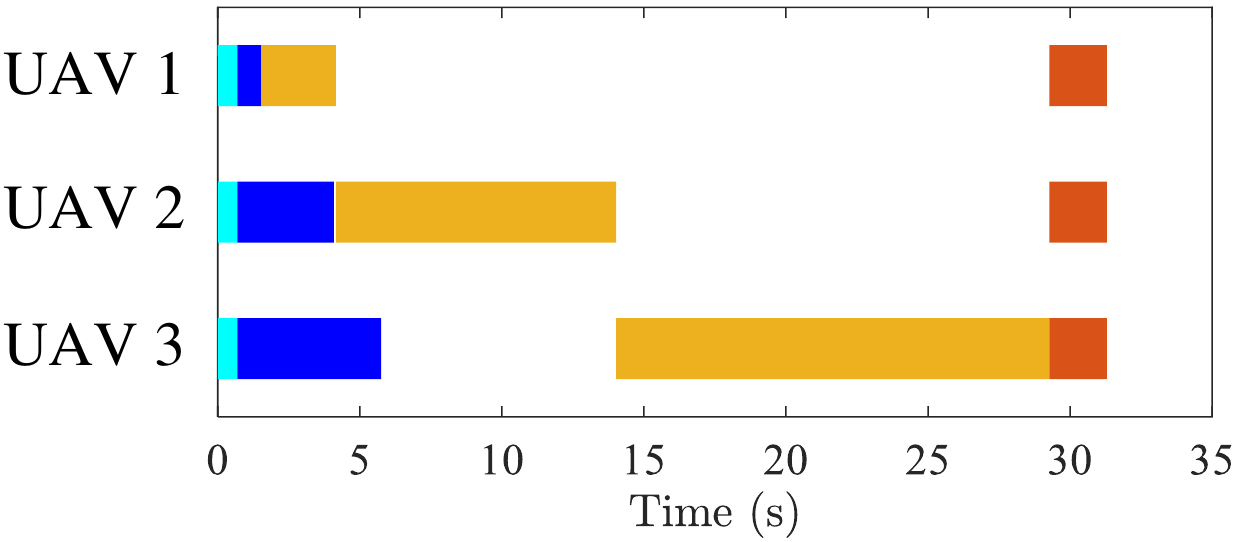}
	}	
	\caption{Illustration of mission completion time of the UAVs.}
	\label{figure4}
\end{figure*}
\subsection{Illustration of the S\&T Process}
\begin{figure}[t]
	\centering
	\setlength{\abovecaptionskip}{0.cm}
	\includegraphics[width=8.5cm]{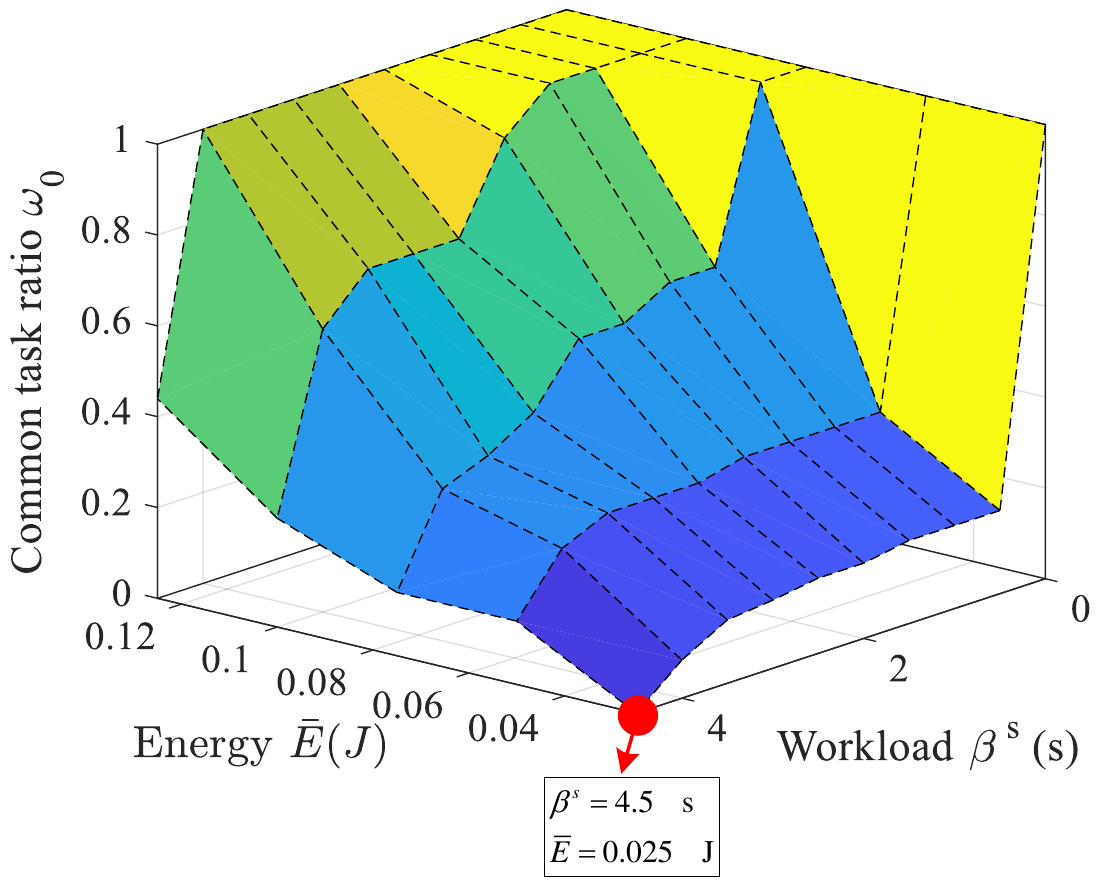}
	\caption{The relationship among workload, energy, and common task ratio.}
	\label{figure7}
\end{figure}

The S\&T process of each UAV under the proposed scheme is illustrated in Fig.~\ref{figure4} for different workload $\beta^s$ and energy budget $\bar E$, where the corresponding sensing time and transmission time are shown in the same color as in Fig.~\ref{figure2}. It can be seen from Figs.~\ref{figure4a} and \ref{figure4c} that when $\bar{E}$ is sufficiently large, the optimal common task ratio $\omega_0 = 0$, which verifies the analysis in Proposition \ref{SufficientConditions2}. In addition, it can be observed from Figs.~\ref{figure4b} and \ref{figure4d} that, as $\beta^s$ increases, the optimal common task ratio $\omega_0$ decreases since more overlapped sensing tasks will lead to a longer sensing time. The relationship among the common task ratio $\omega_0$, the workload $\beta^s$, and the energy budget $\bar E$ is presented in Fig. \ref{figure7}. When the workload $\beta^s$ approaches zero, the optimal $\omega_0 =1$, which implies that each of the UAVs will perform the entire mission; this conforms well to the analysis in Proposition \ref{TsEqual0}. Besides, it can be seen that the optimal common task ratio is higher when $\beta^s$ is small and $\bar E$ is large. Also, it can be observed that, there is no need to perform overlapped sensing (i.e., $\omega_0 = 0$) when $\beta^s$ exceeds 4.5 s and $\bar E$ is below 0.025 J in the considered scenario.

\begin{figure*}[t]
	\centering
	\setlength{\abovecaptionskip}{0.cm}
	\subfigure[$T$ versus workload $\beta^s$ with $\bar {E} = 0.2$ J.]
	{	
		\label{figure5a}
		\includegraphics[width=6.9cm]{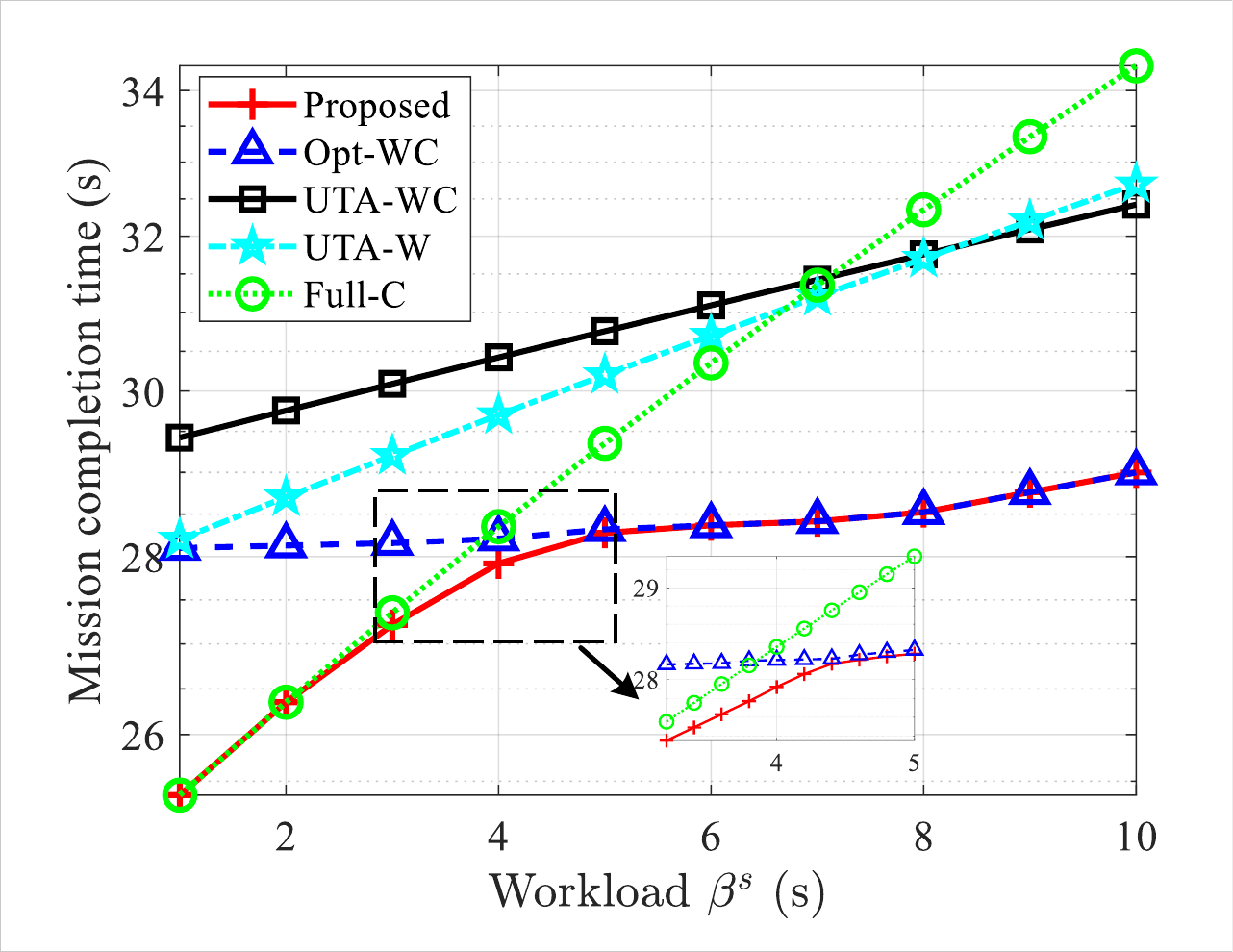}
	}\hspace{6mm}
	\subfigure[$T$ versus workload $\beta^s$ with $\bar {E} = 1$ J.]
	{	
		\label{figure5b}
		\includegraphics[width=6.9cm]{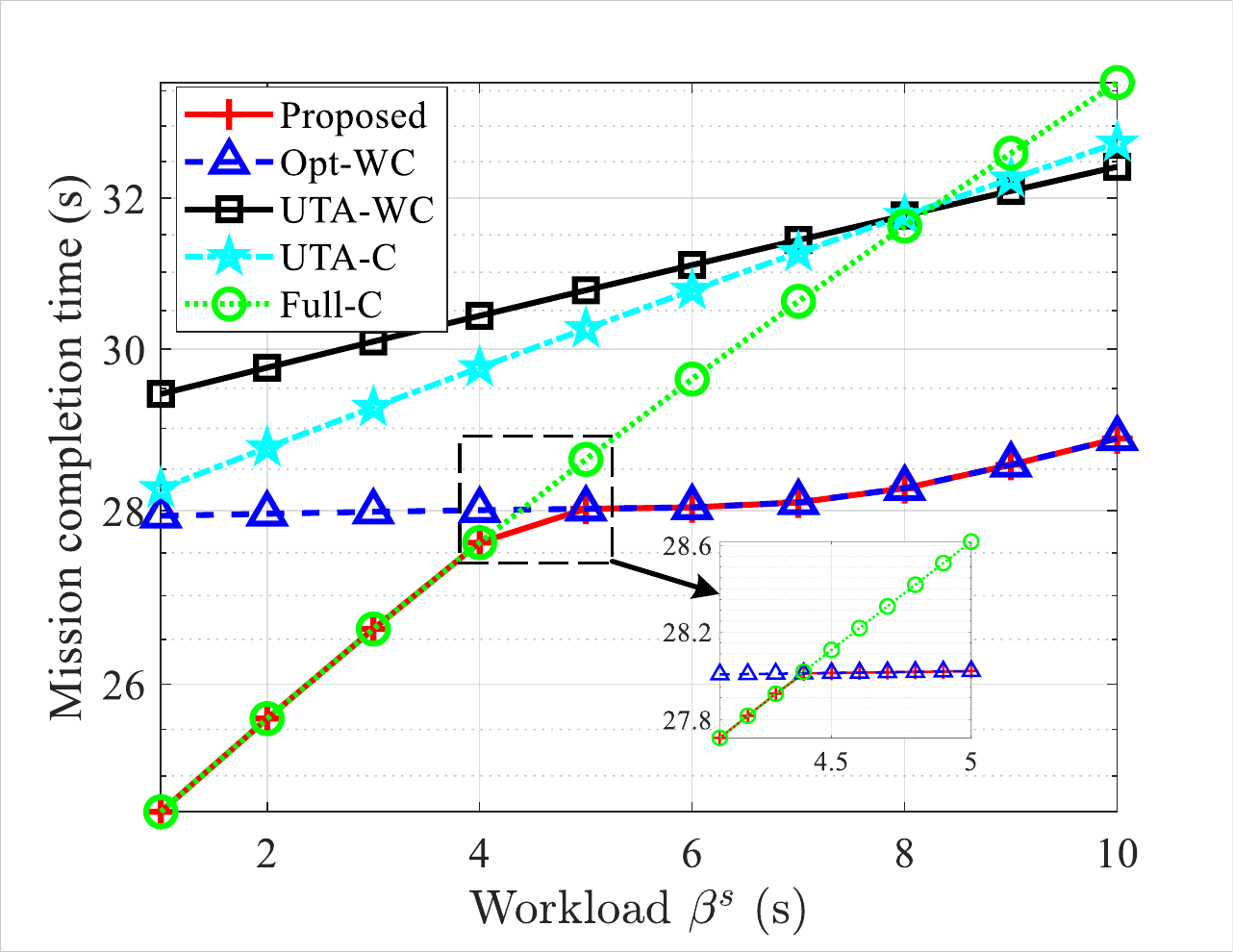}
	}	
	\caption{Mission completion time $T$ comparison under different workloads $\beta^s$ and energy budgets  $\bar {E}$.}
	\label{figure5}
\end{figure*}
\subsection{Task Completion Time Versus Sensing Time} 
The mission completion time $T$ achieved by different schemes are compared in Fig.~\ref{figure5} under different workload $\beta^s$ and energy budget $\bar {E}$. Specifically, it can be seen that, the Opt-WC scheme and the Full-C scheme lead to 14.3\% and 18.6\% longer mission completion time as compared to the proposed scheme, respectively. Besides, it can be observed that the reduction of the mission completion time achieved by our proposed scheme over the Full-C scheme increases as the workload $\beta^s$ increases, while that over the Opt-WC scheme decreases as the workload increases. The main reason is that when the time cost of acquiring sensory data (proportional to $\beta^s$) is small, the improvement of transmission rate due to overlapped sensing is more significant as compared to the corresponding sensing time increment, thereby leading to a significant reduction of the overall mission completion time compared to the Opt-WC scheme. On the other hand, when the time cost of acquiring sensory data is high, compared to the Full-C scheme, the main reason for the reduction in task completion time is due to the proper optimization of the sensing task allocation ratio by the proposed algorithm. This observation conforms well to our analysis in Proposition \ref{TsEqual0}. 

Moreover, it is worth noting that in Fig.~\ref{figure5b}, with energy budget $\bar E = 1$ J, the mission completion time of the proposed algorithm partially coincide with those of the "Opt-WC" and "Full-C" schemes. In Fig.~\ref{figure5b}, the UAVs tend to perform all tasks together when the workload is less than $4.3$ s, since the cooperative data transmission can reduce the total mission completion time. As a result, the corresponding performance is very close to that of the "Full-C" scheme. On the other hand, when the workload exceeds $4.5$ s, the time reduction brought by cooperative transmission cannot compensate for the extra sensing time caused by overlapped sensing. In this case, the UAVs tend to perform the sensing tasks independently, and hence the corresponding performance approaches to that of the "Opt-WC" scheme. Actually, this indicates that the proposed scheme can achieve a good trade-off between transmission and sensing in different settings.

\begin{figure*}[t]
	\centering
	\setlength{\abovecaptionskip}{0.cm}
	\subfigure[$T$ versus $\bar E$.]
	{	
		\label{figure6a}
		\includegraphics[width=6.9cm]{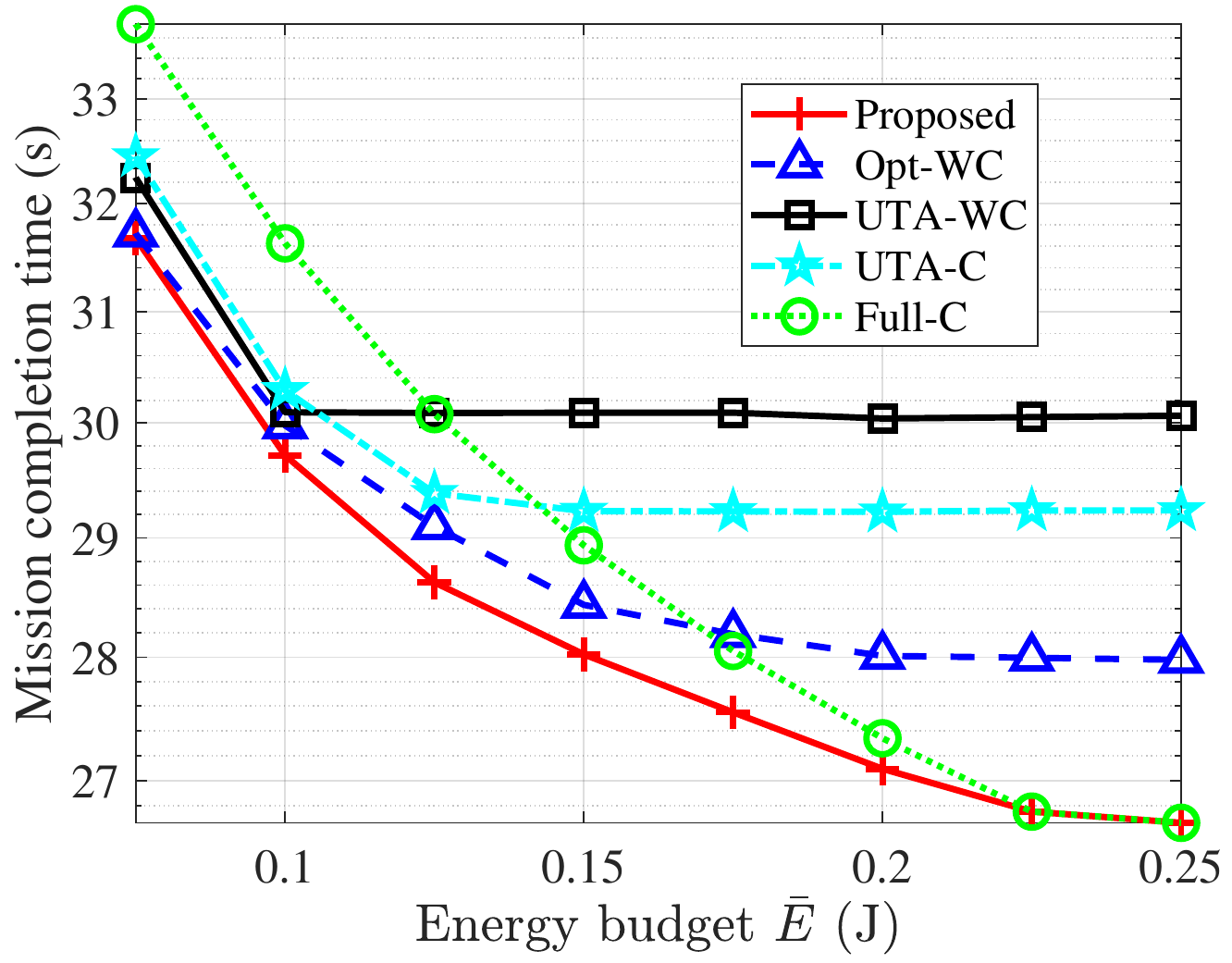}
	}\hspace{6mm}
	\subfigure[$T$ versus $p^{\max}$.]
	{	
		\label{figure6b}
		\includegraphics[width=6.9cm]{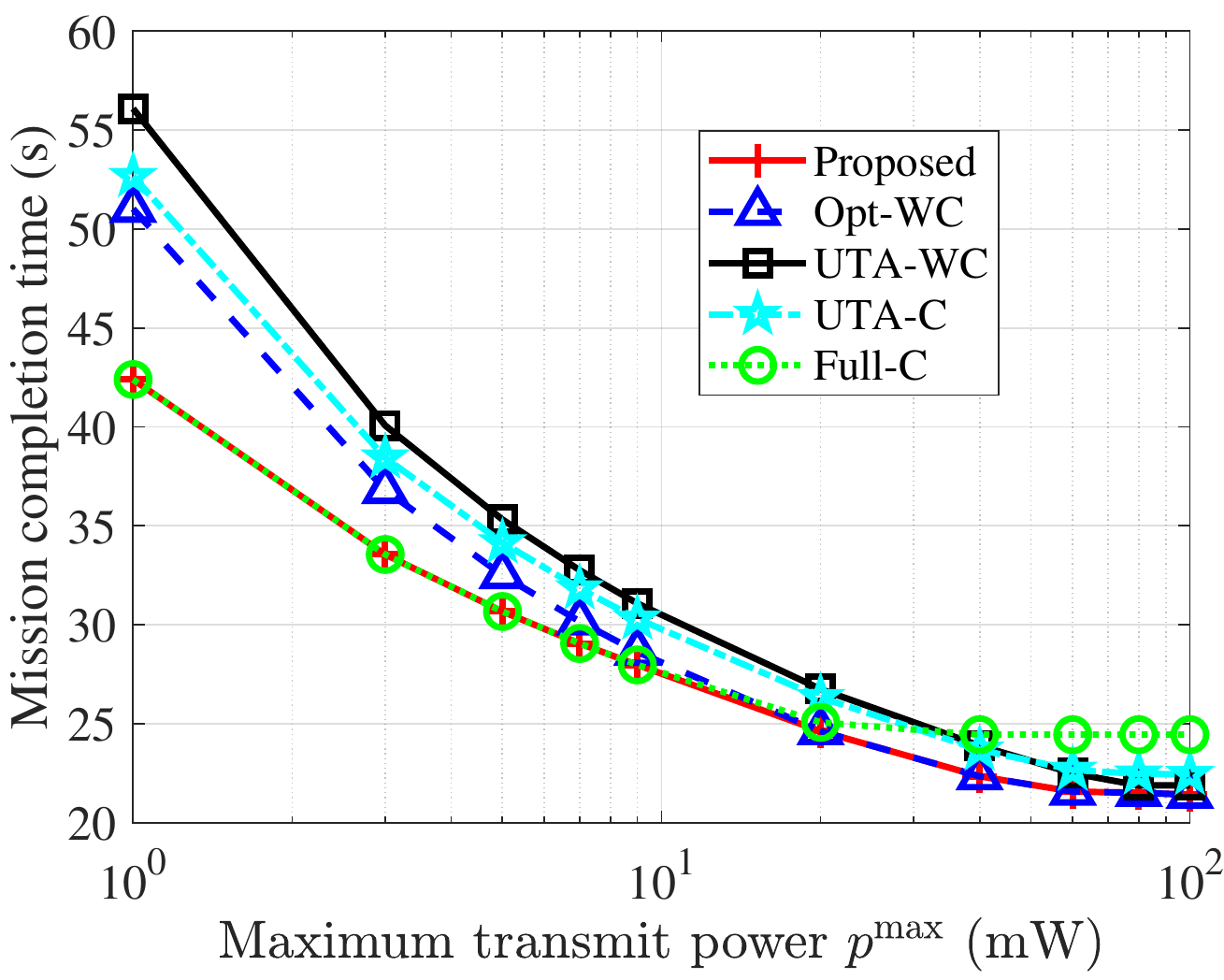}
	}	
	\caption{Mission completion time $T$ comparison under different energy budget $\bar E$ and transmit power $p^{\max}$.}
	\label{figure6}
\end{figure*}
\subsection{Task Completion Time Versus Transmit Power and Energy Budget}
The mission completion time is compared in Fig.~\ref{figure6} for different energy budget $\bar {E}$ and maximum transmit power $p^{\max}$. As shown in Fig.~\ref{figure6}, under different energy budgets $\bar E$ and different maximum transmit power $p^{\max}$, the Opt-WC scheme and the Full-C scheme lead to 20.4\% and 14.1\% longer mission completion time as compared to the proposed scheme, respectively. It is worth noting that the mission completion time of our proposed method is significantly lower compared to the Full-C scheme when the energy budget is small or the maximum transmit power is large. The main reason for this improvement is the following. The ratio of overlapped sensing in Full-C is too large and thus leads to a significant increment in sensing time which cannot be compensated by collaborative transmission. In contrast, the proposed scheme can take the optimal ratio of overlapped sensing to achieve a better tradeoff between transmission time and sensing time, thereby leading to a reduction of the overall mission completion time. On the other hand, the mission completion time of our proposed method is reduced significantly compared to the Opt-WC scheme when the energy budget is larger or the maximum transmit power is smaller. The main reason is that the gain in data transmission rate due to overlapped sensing will be more pronounced when the given energy budget $ \bar{E}$ is sufficient.

\begin{figure*}[t]
	\centering
	\setlength{\abovecaptionskip}{0.cm}
	\subfigure[Mission completion time versus sensing time.]
	{	
		\label{figure8a}
		\includegraphics[width=6.9cm]{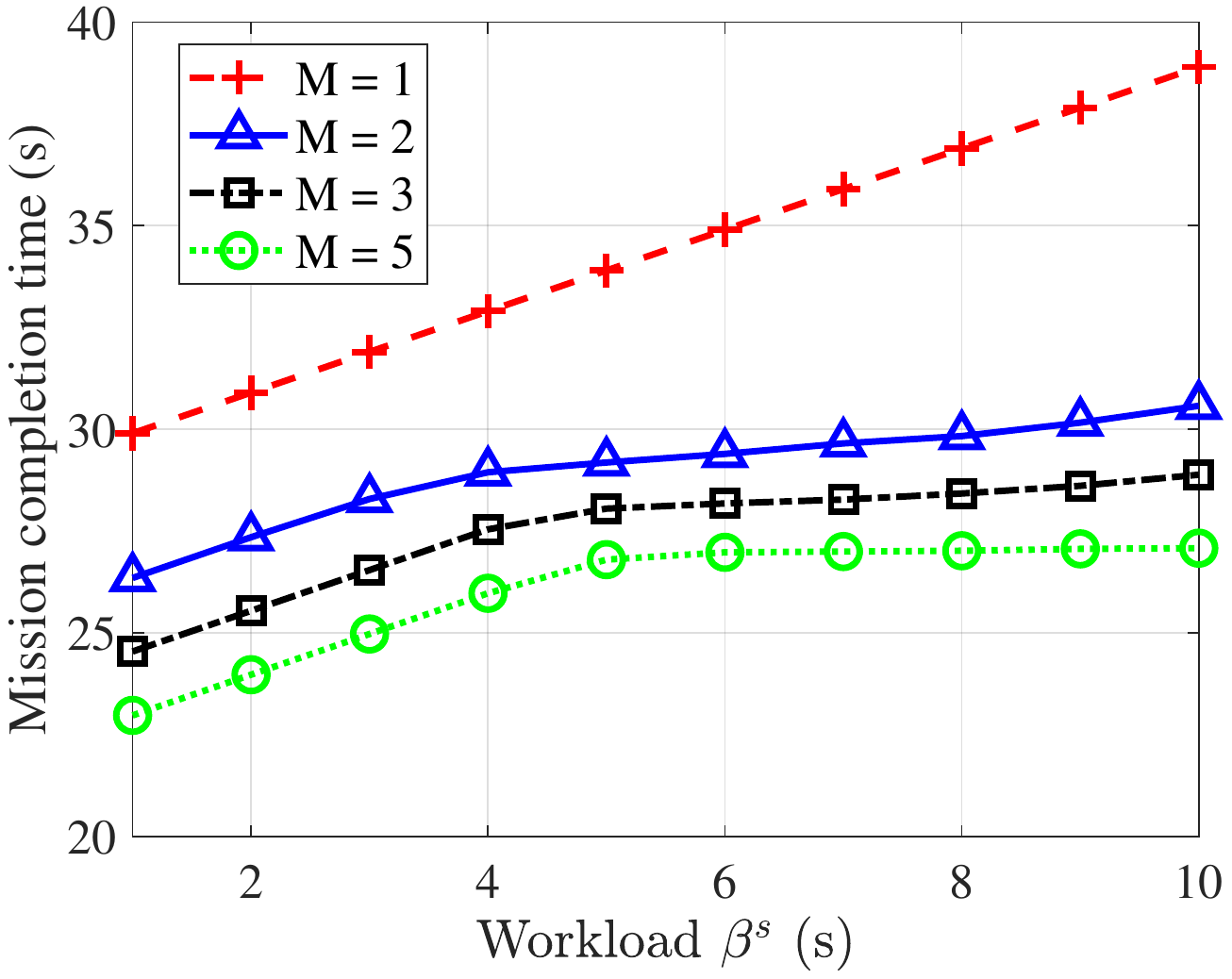}
	}\hspace{6mm}
	\subfigure[Mission completion time versus energy budget.]
	{	
		\label{figure8b}
		\includegraphics[width=6.9cm]{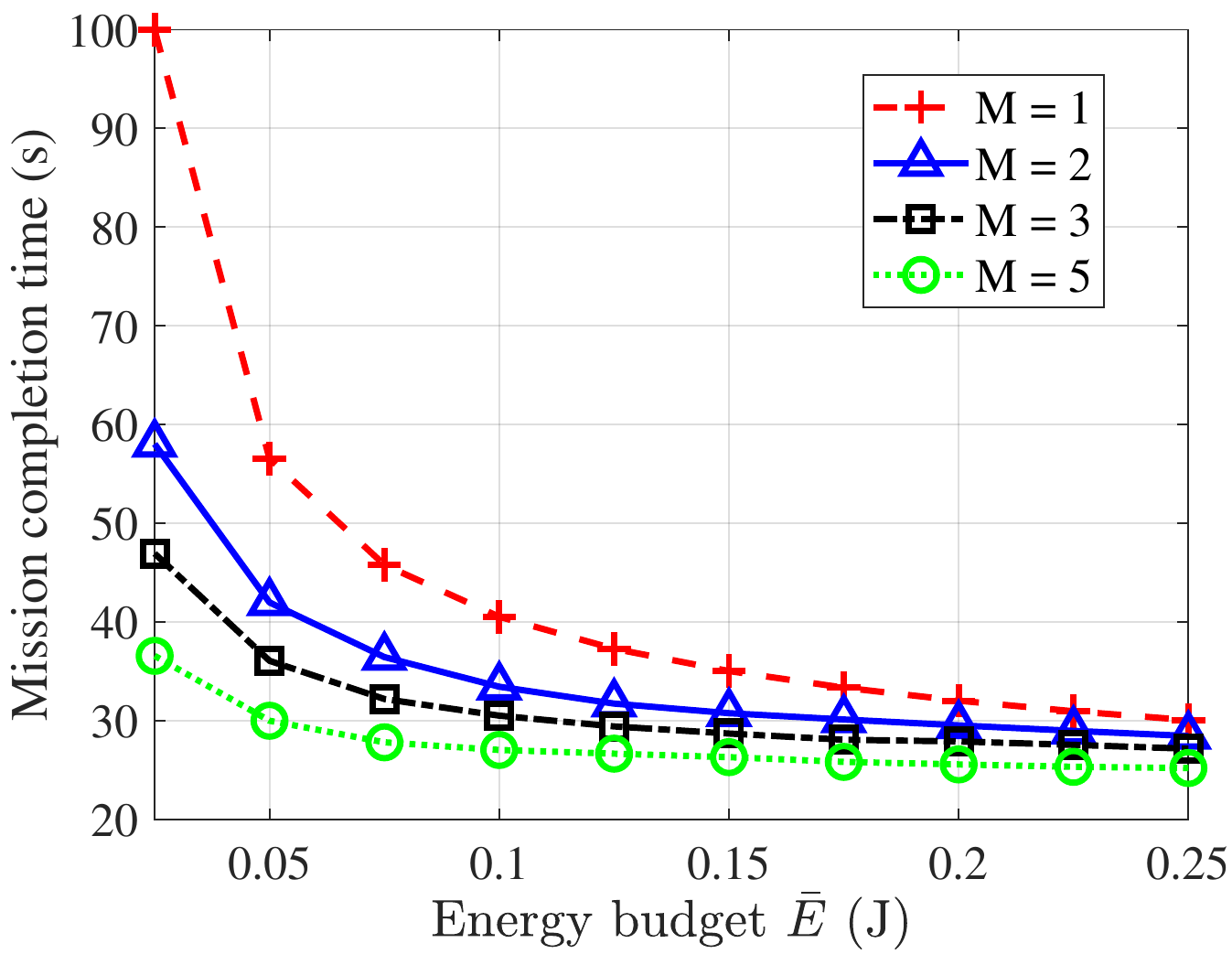}
	}	
	\caption{Mission completion time comparison under different numbers of UAVs.}
	\label{figure8}
\end{figure*}

\subsection{Task Completion Time Versus UAV number}
In Fig.~\ref{figure8a}, it can be seen that the mission completion time decreases when the number of UAVs increases. In addition, it can be observed that as compared to the single-UAV scenario ($M$ = 1), the proposed algorithm under multi-UAV scenarios can bring a greater reduction of the mission completion time, especially when the workload is large. Moreover, Fig.~\ref{figure8b} shows that under a smaller energy budget, the reduction in mission completion time is more pronounced as the number of UAVs increases. The main reason is that when the energy budget is large, the mission completion time is mainly determined by the workload.

\section{Conclusions and Future Works}

\label{Conclusion}
In this paper, the mission completion time minimization problem for multi-UAV S\&T systems is studied and a novel multi-UAV cooperative S\&T scheme with overlapped sensing is proposed. A necessary condition for overlapped sensing is derived. For the cases of overlapped sensing, an MO-based algorithm is proposed by decoupling the transmit time and power variables. For the degenerated case of non-overlapped sensing, by deriving the optimal transmission time in closed-form, a double-loop binary search algorithm is proposed to find the optimal solution. Finally, simulation results demonstrate that our proposed scheme achieves a significantly shorter mission completion time over the benchmark schemes and that the common task ratio plays an important role in optimizing S\&T time. 

The study of clustering-based solutions when the UAVs do not have overlapped field of view and the extension to more sophisticated scenarios with imperfect synchronization are both worthwhile future works.

\normalsize 
\section*{Appendix A: \textsc{Proof of Lemma \ref{EqualProblem1}}}
\label{EqualProblemP1}

\par
In the following, we first prove that at the optimal solution of (P1), $T_1^s + T_1^n \ge T_2^s$ always holds by contradiction. It is assumed that there exists an optimal solution of (P1) which satisfies $T_1^s + T_1^n < T_2^s$, the corresponding task ratio and the energy consumption for independent transmission of UAV 1 (UAV 2) are denoted by $\omega_1^*$ and $E_1^*$ ($\omega_2^*$ and $E_2^*$), respectively. Since $T_1^s + T_1^n < T_2^s$, there always exists a variable $\Delta \omega > 0$, making the following condition holds:
\begin{equation}\label{Lemma_proof_condition}
\left\{ {\begin{array}{*{20}{c}}
	{\left( {\omega _1^* + \Delta {\omega }} \right)  {\beta^s} + \left( {\omega _1^* + \Delta {\omega }} \right)  C  t_1^{n*} = T_2^s}\\
	{\left( {\omega _1^* + \Delta {\omega }} \right)  C  t_1^{n*}  \left( {{2^{\frac{1}{{B  t_1^{n*}}}}} - 1} \right) = {E_1^*}}
	\end{array}} \right.,
\end{equation}
where $t_1^{n*}$ is the optimal transmission time of UAV 1 with task ratio $\omega_1^* + \Delta \omega$ under the energy consumption $E_1^*$ for independent transmission. It can be readily proved $t_1^{n*} > t_1^n$. Accordingly, if $\omega_1^*$ is increased by $\Delta \omega$ and $\omega_2^*$ is decreased by $\Delta \omega$, the transmission time of UAV 2 can be further reduced due to the decreased amount of tasks, thus reducing the total mission completion time $T$. Therefore, $T$ can be reduced by increasing $\omega_1^*$ until $T_1^s + T_1^n = T_2^s$.
In a similar way, we can readily prove that at the optimal solution of (P1), the following conditions hold:
\begin{equation}\label{OptimalSolutionCondition1}
T_m^{\max} + T_m^n \ge T_{m+1}^s, \forall m \in {\cal{M}} \backslash \{M\},
\end{equation}
where $T_1^{\max} = T_1^s$ and $T_m^{\max} = T_{m-1}^{\max} + T_{m-1}^n$. In (\ref{OptimalSolutionCondition1}), $T_m^{\max}$ represents the time instant when UAV $m$ can start data transmission. Then, (P1) can be rewritten as 
\begin{alignat}{2}\label{P1.1Prove}
(\rm{P1.1}): \quad & \mathop {\min }\limits_{{\bm{\omega}} ,\{ p_m^c\} ,\{ p_m^n\} }  T_1^s + \sum\nolimits_{m = 1}^M {T_m^n}  + {T^c} , & \\ 
\mbox{s.t.}\quad
& (\ref{P1-a})-(\ref{P1-e}), & \nonumber \\
& T_m^{\max} + T_m^n \ge T_{m+1}^s,\forall m \in  {\cal{M}} \backslash \{M\}.  & \tag{\ref{P1.1Prove}a}
\end{alignat} 

\section*{Appendix B: \textsc{Proof of Proposition \ref{SufficientConditions}}}
	
If the energy budget in (\ref{P1-a}) satisfies strict inequality for UAV $m$, $p_{m}^c = p^{\max}$ and $p_{m}^n = \min \left( {\frac{1}{{{{\gamma _m}}}}{{{2^{\frac{{C  {\omega _m}}}{{ B  ({T^s_{m+1} - T_m^{\max}})}}}} - 1}},{p^{\max }}} \right)$. Proposition \ref{SufficientConditions} obviously holds. In the following, we will prove that Proposition \ref{SufficientConditions} holds when there are parts of constraints in (\ref{P1-a}) satisfying equality constraints. For ease of analysis, (\ref{P1.1}a) is ignored first to analyze the relationship between the objective function of (P1.1) and the independent transmit power $p_m^n$ of UAV $m$. For given task allocation ratios ${\bm{\omega}}$, when $E_m = \bar {E}$, multiplying both sides of constraint (\ref{P1-a}) with $\gamma_ m$ yields
\begin{equation}\label{EnergyEquation}
\frac{{\omega_0  C  p_m^c  {\gamma _m}}}{{{{\log }_2}(1 + \sum\nolimits_{m = 1}^M {p_m^c{\gamma _m}} )}} + \frac{{{\omega _m}  C  p_m^n  {\gamma _m}}}{{{{\log }_2}(1 + p_m^n{\gamma _m})}} = {{\hat{E}}_m},
\end{equation} 
where ${\hat{E}_{m}} = \bar {E}  \gamma_m$. If $E_m < \bar {E}$, there must exist a constant parameter ${\bar {E}_{m}}$ making the following condition holds:
\begin{equation}
	\frac{{{\omega _0}  C  p^{\max}}}{{{{\log }_2}\left( {1 + \sum\nolimits_{m = 1}^M {p_m^c{\gamma _m}} } \right)}} + \frac{{{\omega _m}  C  p^{\max}}}{{{{\log }_2}\left( {1 + p^{\max}{\gamma _m}} \right)}} = {\bar {E}_{m}},
\end{equation}
Then, by accumulating the corresponding energy consumption of each UAV, the total energy consumption of all UAVs is given by
\begin{equation}\label{EnergySumEquation}
		\frac{{{\omega _0}  \sum\nolimits_{m = 1}^M {p_m^c{\gamma _m}} }}{{{{\log }_2}(1 + \sum\nolimits_{m = 1}^M {p_m^c{\gamma _m}} )}} + \sum\nolimits_{m = 1}^M {\frac{{{\omega _m}  p_m^n  {\gamma _m}}}{{{{\log }_2}(1 + p_m^n{\gamma _m})}}}  = \sum\nolimits_{m = 1}^M {{{\bar {E}}_m}} .
\end{equation}
Then, (\ref{EnergySumEquation}) can be transformed into
\begin{equation}\label{TransformedEquationEnergy}
		\frac{{\sum\nolimits_{m = 1}^M {\tilde p_m^c} }}{{{{\log }_2}(1 + \sum\nolimits_{m = 1}^M {\tilde {p}_m^c} )}} = {\sum\nolimits_{m = 1}^M {{{\tilde E}_m}} 
			- \sum\nolimits_{m = 1}^M {{{\tilde \omega }_m}  \frac{{\tilde p_m^n}}{{{{\log }_2}(1 + \tilde p_m^n)}}} } ,
\end{equation}
where $\tilde p_m^c = p_m^c{\gamma _m}$, $\tilde p_m^n = p_m^n{\gamma _m}$, ${{\tilde E}_m} = {{{\bar E_m}} \mathord{\left/{\vphantom {{{\bar E_m}} {\left({\omega _0  C}\right)}}} \right.\kern-\nulldelimiterspace} {\left({\omega _0  C}\right)}}$, and ${{\tilde \omega }_m}{{ = }}{{{\omega _m}} \mathord{\left/{\vphantom {{{\omega _m}} {{\omega _0}}}} \right.\kern-\nulldelimiterspace} {{\omega _0}}}$.
In the following, we analyze the relationship between the mission completion time $T$ and $\tilde p_m^n$, the relationship between $T$ and $p_m^n$ can be further obtained, Letting $x = \frac{1}{{{{\log }_2}(1 + \sum\nolimits_{m = 1}^M {\tilde p_m^c} )}}$ and $a = \sum\nolimits_{m = 1}^M {{{\tilde E}_m}}  - \sum\nolimits_{m = 1}^M {{{\tilde \omega }_m}  \frac{{\tilde p_m^n}}{{{{\log }_2}(1 + \tilde p_m^n)}}} $,  (\ref{TransformedEquationEnergy}) is rewritten as 
\begin{equation}\label{a_function_x}
a(x) = x  \left( {{2^{\frac{1}{x}}} - 1} \right).
\end{equation}
By taking the differentiation on both sides of (\ref{a_function_x}), we have
\begin{equation}
\frac{{\partial x}(a)}{{\partial a}} = \frac{1}{{{2^{\frac{1}{x}}}\left( {1 - \frac{{\ln 2}}{x}} \right) - 1}}.
\end{equation}
\par
Define $f(x) = \frac{1}{{{2^{\frac{1}{x}}}\left( {1 - \frac{{\ln 2}}{x}} \right) - 1}}$. $f(x)$ is a monotonically increasing function about $x$ since $f'(x) \ge 0$. In the following, it will be proved that at the optimal solution of (P1.1), $x \le \frac{1}{{{{\log }_2}(1 + p_m^u{\gamma _m})}}$ by contradiction. Assume that there exists an optimal solution when $x > \frac{1}{{{{\log }_2}(1 + p_m^u{\gamma _m})}}$, $m \in {\cal{M}}$. If ${\bar{E}_{m}} < \bar {E}$, we have
\begin{equation}\label{ParitialCOndition}
\begin{aligned}
\frac{{\partial {{T}}}}{{\partial \tilde p_m^n}} &= \frac{{\partial x(a)}}{{\partial a}}  \frac{{\partial a}}{{\partial \tilde p_m^n}} + {{\tilde \omega }_m}{\frac{{\partial \left( {\frac{1}{{{{\log }_2}(1 + \tilde p_m^n)}}} \right)}}{{\partial \tilde p_m^n}} }\\
&= \frac{{{{\tilde \omega }_m}}}{{1 - {2^{\frac{1}{x}}}\left( {1 - \frac{{\ln 2}}{x}} \right)}}  \frac{{\partial \left( {\frac{{\tilde p_m^n}}{{{{\log }_2}(1 + \tilde p_m^n)}}} \right)}}{{\partial \tilde p_m^n}} + {{\tilde \omega }_m}  \frac{{\partial \left( {\frac{1}{{{{\log }_2}(1 + \tilde p_m^n)}}} \right)}}{{\partial \tilde p_m^n}}\\
&\overset{(b)}{\ge} \frac{{{{\tilde \omega }_m}  \frac{{\partial \left( {\frac{{\tilde p_m^n}}{{{{\log }_2}(1 + \tilde p_m^n)}}} \right)}}{{\partial \tilde p_m^n}}}}{{1 - (1 + \tilde p_m^n)(1 -   {{\log }_2}(1 + \tilde p_m^n)\ln 2)}} + {{\tilde \omega }_m}  \frac{{\partial \left( {\frac{1}{{{{\log }_2}(1 + \tilde p_m^n)}}} \right)}}{{\partial \tilde p_m^n}}\\
&{{ = }}\frac{{{{\tilde \omega }_m}}}{{\left( {{{1 + }}\tilde p_m^n} \right)  {{\log }_2}^{{2}}(1 + \tilde p_m^n)\ln 2}}  \left( {\frac{{\left( {{{1 + }}\tilde p_m^n} \right){{\log }_2}(1 + \tilde p_m^n)\ln 2 - \tilde p_m^n}}{{1 - (1 + \tilde p_m^n)(1 -   {{\log }_2}(1 + \tilde p_m^n)\ln 2)}} - 1} \right) \\
&= 0.
\end{aligned}
\end{equation}	
In (\ref{ParitialCOndition}), ($b$) holds since $x \ge \frac{1}{{{{\log }_2}(1 + p_m^u{\gamma _m})}}$. Accordingly, the objective function of (P1.1) is an increasing function of $\tilde p_m^n$ when $x > \frac{1}{{{{\log }_2}(1 + p_m^u{\gamma _m})}}$. Hence, the mission completion time $T$ can always be reduced by decreasing $\tilde p_m^n$ until $x = \frac{1}{{{{\log }_2}(1 + p_m^u{\gamma _m})}}$. Therefore, there is no optimal solution when $x > \frac{1}{{{{\log }_2}(1 + p_m^u{\gamma _m})}}$, $m \in {\cal{M}}$. Similarly, we can readily prove that $\frac{{\partial {{T}}}}{{\partial \tilde p_m^n}} \le 0$ when $x < \frac{1}{{{{\log }_2}(1 + p_m^u{\gamma _m})}}$. Therefore, $p_M^n = \min(\sum\nolimits_{m = 1}^M {p_m^c{\gamma _m}}, p^{\max})$.

Considering the constraints in (\ref{P1.1}a), $p_m^n \le \bar {p}_m^n = \frac{{{2^{\frac{{C  {\omega _m}}}{{ B  ({T^s_{m+1} - T_m^{\max}})}}}} - 1}}{{{\gamma _m}}}, m \in  {\cal{M}} \backslash \{M\}$. By combing the above results, the optimal transmit power $p_m^n$ of UAV $m$ can be given by
\begin{equation}
	p_m^n = \left\{ {\begin{array}{*{20}{c}}
			{\min(\sum\nolimits_{m = 1}^M {p_m^c{\gamma _m}}, p^{\max}),}&{m = M}\\
			{\min( \sum\nolimits_{m = 1}^M {p_m^c{\gamma _m}}, \min ( {\bar{p}_m^n,{p^{\max }}} )) ,}&{{\rm{otherwise}}}
	\end{array}}. \right.
\end{equation}

\section*{Appendix C: \textsc{Proof of Corollary \ref{OptimalCooperativeTime}}}

In the following, Corollary \ref{OptimalCooperativeTime} will be proved for two composite cases.
\begin{itemize}
	\item Case A: All constraints in (\ref{P1-a}) satisfy with strict inequality;
	\item Case B: Parts of constraints or no constraints in (\ref{P1-a}) satisfy with strict inequality;
\end{itemize}
First, for Case A, all constraints in (\ref{P1-a}) satisfies with strict inequality at the optimal solution of (P1), if the energy budget constraint in (\ref{P1-a}) satisfies strict inequality for UAV $m$, we have $p_{m}^c = p^{\max}$ and $p_{m}^n = p^{\max}$. Since $\sum\nolimits_{m = 1}^M {{p^{\max }}{\gamma _m}}  \ge {p^{\max }}{\gamma _m}$, $\forall m \in {\cal{M}}$, it follows that ${t^{c*}} \le {t^{n*}_m}$. Corollary \ref{OptimalCooperativeTime} obviously holds.
\par
For Case B, according to proof of Proposition \ref{SufficientConditions} in Appendix B, the objective function of (P1.1) is an increasing function of $\tilde p_m^n$ when $\frac{1}{{{{\log }_2}(1 + \sum\nolimits_{m = 1}^M {\tilde p_m^c} )}} > \frac{1}{{{{\log }_2}(1 + p_m^u{\gamma _m})}}$. Accordingly, at the optimal solution of (P1.1), $\frac{1}{{{{\log }_2}(1 + \sum\nolimits_{m = 1}^M {\tilde p_m^c} )}} \le \frac{1}{{{{\log }_2}(1 + p_m^u{\gamma _m})}}$, $m \in {\cal{M}}$. Therefore, ${t^{c*}} \le {t^{n*}_m}$, $\forall m \in {\cal{M}}$. This completes the proof.

\section*{Appendix D: \textsc{Proof of Proposition \ref{SufficientConditions2}}}
In the following, it will be proved that there is no need for overlapped sensing if $x \le p^{\max} \gamma_M$ by contradiction. First, it will be proved that, the minimum total transmission time is achieved when $\omega_0 = 1$.
Since $f(x) = \frac{1}{{{\log }_2}(1 + x)}$ is convex about $x$, for any given task allocation $\{\omega_1,\cdots,\omega_M\}$ and transmit power ${\bm{p}}^n$, the following condition holds:
\begin{equation}\label{CooperativeTransmissionW0zero}
	\underbrace{\sum\nolimits_{m = 1}^M \frac{\omega_m C}{{ B{{\log }_2}(1 + p_m^n  {\gamma _m})}}}_{\text{Independent transmission time}} \overset{(c)}{\ge}  \underbrace{\frac{\Omega C}{{ B{{\log }_2}(1 + \sum\nolimits_{m = 1}^M {\frac{\omega_m}{\Omega} p_m^n  {\gamma _m}})}}}_{\text{Cooperative transmission time}},
\end{equation}
where $\Omega = \sum\nolimits_{m = 1}^M \omega_m$ and ($c$) holds due to the convexity of $f(x)$. Moreover, condition ($c$) in (\ref{CooperativeTransmissionW0zero}) becomes active when $p_m^n  {\gamma _m} =  p_{m+1}^n  {\gamma _{m+1}}, \forall m \in {\cal{M}} \backslash \{M\}$. It can be readily proved that the transmission time for cooperative transmission is $t^c \le t^n_m$ when the energy budget is enough. In (\ref{CooperativeTransmissionW0zero}), ${\frac{\omega_m}{\Omega} p_m^n }$ and $\sum\nolimits_{m = 1}^M {\frac{\omega_m}{\Omega} p_m^n  {\gamma _m}}$ respectively represent the transmit power for cooperative transmission of UAV $m$ and the equivalent SNR for cooperative data transmission, where the energy consumption of UAV $m$ for this cooperative transmission satisfies $\frac{({\frac{\omega_m}{\Omega} p_m^n}) \Omega C}{{B{{\log }_2}(1 + \sum\nolimits_{m = 1}^M {\frac{\omega_m}{\Omega} p_m^n {\gamma _m}})}} \le \frac{p_m^n \omega_m C}{{ B{{\log }_2}(1 + p_m^n  {\gamma _m})}} \le \bar {E}$. Hence, both the transmission time and the energy consumption of cooperative transmission are no more than those of independent transmission, i.e., the minimum total transmission time is achieved when $\omega_0 = 1$. 

Then, it will be proved that the total transmission time for cooperative transmission equals to that for independent transmission if the energy budget is not enough, i.e., the solution $x$ of (\ref{CooperativeEquation2}) satisfies $x > p^{\max} \gamma_M$,
\begin{equation}\label{CooperativeEquation2}
	\frac{{  C  x }}{{B{{\log }_2}(1 + x )}} = \bar {E} \sum\nolimits_{m = 1}^M   \gamma_m.
\end{equation} 
Specifically, multiplying both sides of (\ref{P1-a}) with $\gamma_ m$ and accumulating these energy consumption constraints yields
\begin{equation}
	\frac{{  C  \sum\nolimits_{m = 1}^M {p_m^c  {\gamma _m}} }}{{B {{\log }_2}(1 + \sum\nolimits_{m = 1}^M {p_m^c{\gamma _m}} )}} = \bar {E} \sum\nolimits_{m = 1}^M   \gamma_m.
\end{equation}
Since the energy consumption and the transmission times of all UAV are equal for cooperative transmission, the transmit power of each UAV is also equal, i.e., $p_m^c = p_{m+1}^c$, $\forall m \in {\cal{M}} \backslash \{M\}$. In the following, it will be proved that we could always construct another solution with $\omega_0 = 0$ if $\sum\nolimits_{m = 1}^M {p_m^c{\gamma _m}} \le p^{\max} \gamma_M$. Specifically, for individual tasks, the task allocation ${\bm{\omega}}$ and the transmit power ${\bm{p^n}}$ are constructed as follows:
\begin{equation}
	\omega_m = \frac{\gamma_m}{\sum\nolimits_{m = 1}^M \gamma_m}, \quad {p_m^n } = \frac{\sum\nolimits_{m = 1}^M p_m^c \gamma_m}{\gamma_m},
\end{equation}
where ${p_m^n } \le p^{\max}$. In this case, it is not difficult to verify that the total transmission time for individual tasks is equal to that for $\omega_0 = 1$. Also, the energy consumption of UAV $m$ satisfies
\begin{equation}
	\frac{p_m^n \omega_m C}{B {\log_2 (1+ p_m^n \gamma_m)}} = \frac{\sum\nolimits_{m = 1}^M p_m^c  \frac{\gamma_m}{\sum\nolimits_{m = 1}^M \gamma_m} C}{B {\log_2 (1+ \sum\nolimits_{m = 1}^M {p_m^c{\gamma _m}})}} \le \bar {E}.
\end{equation}
Therefore, when the constructed transmit power satisfies ${p_m^n } \le p^{\max}$, $\forall m$, the energy consumption of each UAV for the data transmission of individual tasks is equal to that of the common task with $\omega_0 = 1$. On the other hand, the sensing time with non-overlapped sensing (i.e., $\omega_0 = 0$) is also no more than that with fully overlapped sensing (i.e., $\omega_0 = 1$). Thus, there is no need for overlapped sensing if the obtained SNR of cooperative transmission is large than $p^{\max} \gamma _m$, $\forall m$, i.e., $\sum\nolimits_{m = 1}^M {p_m^c{\gamma _m}} > p^{\max} \gamma _M$.

\section*{Appendix E: \textsc{Proof of Proposition \ref{TsEqual0}}}
\vspace{-2mm}

If $T_s \to 0$, i.e., the sensing time is negligible, the constraints in (\ref{P1.1-a}) always hold. In this case, all the sensory data can be obtained by each UAV without time consuming. It has been proved in Proposition \ref{SufficientConditions2} that the minimum total transmission time is achieved when $\omega_0 = 1$. Since $T_s \to 0$, the total mission completion time is minimized when $\omega_0 = 1$. Hence, it is proved that if $\beta^s \to 0$, the optimal cooperative task ratio $\omega^*_0 = 1$. 
\par
If $\beta^s \to \infty$, the transmission time can be ignored since the mission completion time approximately equals the maximum sensing time of the UAVs. Thus, it is not necessary to repetitively perform the tasks, i.e., $\omega_0^* = 0$. 

\vspace{-2mm}
\section*{Appendix F: \textsc{Proof of Lemma \ref{T_function_w}}}
\vspace{-1mm}
In the following, this lemma will be proved in two cases: 1) $T^s_{m+1} - T^s_m = T_m^n $; 2) $T^s_{m+1} - T^s_m < T_m^n$.
\par
For the first case $T^s_{m+1} - T^s_m = T_m^n $, it is assumed that there exists an optimal solution ${\bm{\omega}}^* = \{\omega_1^*,...,\omega_M^*\}$ of (P3.1) that makes $\frac{\partial {{\cal{T}}(\omega_m)}}{{{\partial \omega _m}}} |_{\omega_m^*} < \frac{{\partial {{\cal{T}}(\omega_m)}}}{{{\partial \omega _{m + 1}}}}|_{\omega_{m+1}^*}$. Let $\omega'_m$ = $\omega^*_m + \Delta \omega$, and $\omega'_{m+1} = \omega_{m+1}^* - \Delta \omega$, then ${\omega' _m}{{\tau}_{m}}({\omega' _m}) + {\omega' _{m + 1}}{T_{m + 1}}({\omega' _{m + 1}}) < {\omega^* _m}{{\tau}_{m}}({\omega^* _m}) + {\omega^* _{m + 1}}{T_{m + 1}}({\omega^* _{m + 1}})$. Hence, the optimal solution ${\bm{\omega}}^*$ can always be improved if $\frac{\partial {{\cal{T}}(\omega_m)}}{{{\partial \omega _m}}} |_{\omega_m^*} < \frac{{\partial {{\cal{T}}(\omega_m)}}}{{{\partial \omega _{m + 1}}}}|_{\omega_{m+1}^*}$. Hence, Lemma \ref{T_function_w} for the first case is proved. 
\par
For the second case $T^s_{m+1} - T^s_m < T_m^n$, if there exists an optimal solution ${\bm{\omega}}^* = \{\omega_1^*,...,\omega_M^*\}$ of (P3.1) that makes $\frac{\partial {{\cal{T}}(\omega_m)}}{{{\partial \omega _m}}} |_{\omega_m^*} < \frac{{\partial {{\cal{T}}(\omega_m)}}}{{{\partial \omega _{m + 1}}}}|_{\omega_{m+1}^*}$ or $\frac{\partial {{\cal{T}}(\omega_m)}}{{{\partial \omega _m}}} |_{\omega_m^*} > \frac{{\partial {{\cal{T}}(\omega_m)}}}{{{\partial \omega _{m + 1}}}}|_{\omega_{m+1}^*}$, similarly, the mission completion time can be improve until $\frac{\partial {{\cal{T}}(\omega_m)}}{{{\partial \omega _m}}} |_{\omega_m^*} = \frac{{\partial {{\cal{T}}(\omega_m)}}}{{{\partial \omega _{m + 1}}}}|_{\omega_{m+1}^*}$. If $\gamma _m \ne \gamma_{m+1}$, $\frac{\partial {{\cal{T}}(\omega_m)}}{{{\partial \omega _m}}} |_{\omega_m^*} = \frac{{\partial {{\cal{T}}(\omega_m)}}}{{{\partial \omega _{m + 1}}}}|_{\omega_{m+1}^*}$ has a unique solution since $\frac{\partial {{\cal{T}}(\omega_m)}}{{{\partial \omega _m}}} $ is a monotonically increasing function w.r.t $\omega_m$.

\vspace{-2mm}
\section*{Appendix G: \textsc{Proof of Lemma \ref{OptimalCondition}}}
\vspace{-1mm}
If constraints in (\ref{P1-a}) satisfy with strict inequality, i.e., $E_m < \bar {E}$, then $p^n_m = p^{\max}$, and $t^n_m = {\frac{1}{{B{{\log }_2}\left( {1 + {p^{\max }}{\gamma _m}} \right)}}}$. If ${\omega _m} \ge \frac{{B{{\log }_2}\left( {1 + {p^{\max }}{\gamma _m}} \right)  {{\bar {E}}}}}{{{p^{\max }}}}$, constraints in (\ref{P1-a}) will satisfy with strict equality, the following condition holds,
\begin{equation}\label{Energy_equation_1}
{\omega _m}  C  t_m^n  \frac{{\left( {{2^{^{\frac{1}{{B  t_m^n}}}}} - 1} \right)}}{{{\gamma _m}}} = {{\bar {E}}}.
\end{equation}
(\ref{Energy_equation_1}) can be transformed into (\ref{Energy_equation_2}).
\begin{equation}\label{Energy_equation_2}
{2^{^{\frac{1}{{B  t_m^n}}}}} - 1 = \frac{{B  {\gamma _m} \cdot{{\bar {E}}}}}{{ {\omega _m}  C}}  \frac{1}{{B  t_m^n}}.
\end{equation}
Let $x = \frac{1}{{B  t_u^{uc}}}$, the following condition holds
\vspace{-2mm}
\begin{equation}
Q  {2^x} = x + Q,
\vspace{-2mm}
\end{equation}
where $Q = {{ {\omega _m}  C} \mathord{\left/
		{\vphantom {{ {\omega _m}  C} {\left( {B  {\gamma _m} {{\bar {E}}}} \right)}}} \right.
		\kern-\nulldelimiterspace} {\left( {B  {\gamma _m} {{\bar {E}}}} \right)}}$, and multiply both sides of (\ref{Energy_equation_2}) by $2^{(-x-Q)}$, the following condition holds:
\vspace{-2mm}
\begin{equation}
- \ln 2Q  {e^{ - \ln 2Q}} = \ln 2  \left( { - x - Q} \right)  {e^{\ln 2  \left( { - x - Q} \right)}}.
\vspace{-2mm}
\end{equation}
The above equation has the form of $y  {e^y} = \eta \left( {\eta  > 0} \right)$. $\ln 2  \left( { - x - Q} \right) <  - \ln 2Q$ as $x>0$. Then, $x =  - \frac{1}{{\ln 2}}W _{-1}\left( { - \ln 2Q  {e^{ - \ln 2Q}}} \right) - Q$. Let $A_m{\omega _m} = \ln 2Q$. Then, $t_m^n$ can be given by
\begin{equation}
t_m^n = \frac{{ - \ln 2}}{{B  \left( {W _{-1}\left( { - A_m{\omega _m}  {e^{ - A_m{\omega _m}}}} \right) + A_m{\omega _m}} \right)}},
\end{equation}
where $A_m = \frac{{C \ln 2  }}{{B  {\gamma _m}  {{\bar {E}}}}}$. Hence, the optimal transmission time $t^n_m$ can be given by
\begin{equation}
{{\tau}_{m}}({\omega _m}) = \left\{ {\begin{array}{*{20}{c}}
	{\frac{{ - \ln 2}}{{B  \left( {W _{-1}\left( { - A_m{\omega _m}  {e^{ - A_m{\omega _m}}}} \right) + A_m{\omega _m}} \right)}},}&{{\omega _m} \ge {{\hat \omega }_m}}\\
	{\frac{1}{{B{{\log }_2}\left( {1 + {p^{\max }}{\gamma _m}} \right)}}}&{{\omega _m} \le {{\hat \omega }_m}}
	\end{array}} \right. .
\end{equation}
\par
Constraints (\ref{P1-b})$-$(\ref{P1-c}) are convex. For the objective function of (P3.1), when ${{\omega _m} < {{\hat \omega }_m}}$, $f_m(\omega_{m}) = \omega_{m}  {{\tau}_{m}}({\omega _m})$ is convex about $\omega _m$ since $f_m(\omega_{m})$ is a linear function of $\omega _m$. In the following, it will be proved that $f_m(\omega_{m}) = \omega_{m}  {{\tau}_{m}}({\omega _m})$ is convex when ${{\omega _m} \ge {{\hat \omega }_m}}$, i.e., 
\begin{equation}
f_m(\omega_{m}) = \frac{{ - {\omega _m}\ln 2}}{{B  \left( {{W_{ - 1}}\left( { - A_m{\omega _m}  {e^{ - A_m{\omega _m}}}} \right) + A_m{\omega _m}} \right)}}.
\end{equation}
For ease of analysis, define $g_m(x)$ as
\begin{equation}
g_m(x) = \frac{{ - x}}{{{W_{ - 1}}\left( { - {A_m}x  {e^{ - {A_m}x}}} \right) + {A_m}x}}.
\end{equation}
Function $f_m(\omega_{m})$ and function $g_m(x)$ share the same concavity and convexity. Specifically, the first derivative of $g_m(x)$ can be given by
\begin{equation}\label{GmDerva1}
g'_m(x) = \frac{{ - {W_{ - 1}}\left( {{x_m}  {e^{{x_m}}}} \right)}}{{\left( {{W_{ - 1}}\left( {{x_m}  {e^{{x_m}}}} \right) - {x_m}} \right)\left( {1 + {W_{ - 1}}\left( {  {x_m}  {e^{  {x_m}}}} \right)} \right)}},
\end{equation}\label{GmDerva2}
where ${x_m} =  - {A_m}x$. Since ${{W_{ - 1}}\left( { x_m  {e^{ x_m}}} \right) - x} < 0$ and ${{W_{ - 1}}\left( { x_m  {e^{ x_m}}} \right)} < 0$, $g'(x) > 0$. The second derivative of $g_m(x)$ can be given by
\begin{equation}
g''_m(x) = \frac{{W_{ - 1}}\left( { - A_mx  e^{ - A_mx}} \right)}{x{\left( {1 + {W_{ - 1}}\left( { - A_mx  e^{ - A_mx}} \right)} \right)^3}} > 0.
\end{equation}
$g_m(x)$ is convex about $x$ and $f_m(\omega_{m})$ is also convex about $\omega_{m}$ when ${{\omega _m} \ge {{\hat \omega }_m}}$.

\section*{Appendix H: \textsc{Proof of Proposition \ref{MonoIncreasing}}}

For a given $\omega_m$, the optimal task ratio $\omega_{m+1}$ of UAV $m+1$ can be obtained by solving the equations $T^s_{m+1} - T^{\max}_m = {\omega^* _m}{{\tau}_{m}}({\omega^* _m})$ and $\frac{{ {\partial {{\cal{T}}(\omega_m)}}}}{{{\partial \omega _m}}}|_{\omega_m^*} = \frac{{\partial {{\cal{T}}(\omega_{m+1})}}}{{\partial{\omega _{m + 1}}}}|_{\omega_{m+1}^*}$ via binary search, since both $\frac{{ {\partial {{\cal{T}}(\omega_m)}}}}{{{\partial \omega _m}}}$ and $\omega_{m}  {{\tau}_{m}}({\omega _m})$ monotonically increases w.r.t $\omega _m$ (c.f. (\ref{GmDerva1}) in Appendix G). As a result, the task ratio $\omega_{m+1}$ of UAV $m+1$ admitting Lemma \ref{OptimalCondition} will monotonically increase as $\omega_{m}$ increases. Hence, when $\omega_1$ increases, the task ratios of other UAVs $\{\omega_m\}_{m=2}^M$ admitting Lemma \ref{OptimalCondition} will increase. As a result, the corresponding mission completion time will also increase monotonically with $\omega_1$.

\section*{Appendix I: \textsc{Proof of Proposition \ref{NonCooperativeOmega_Relation}}}
\vspace{-1mm}
For variable set ${\cal{P}}$, $\omega_{m} > {\hat \omega}_{m}$, $\forall m \in {\cal{M}}$, and $T_{m+1}^s - T_{m}^{\max} \le T_m^n$, $\forall m \in {\cal{M}} \backslash \{M\}$. Construct Lagrange function $L\left( {{\omega _1},...,{\omega _M}}{,\mu } \right) = {\omega _M}  {\beta^s} + \sum\nolimits_{m = 1}^M {{\omega _m}  C  {{\tau}_{m}}({\omega _m})}  + \mu \left( {\sum\nolimits_{m = 1}^M {{\omega _m}}  - 1} \right)$, where $\mu$ is Lagrange dual factor. The Lagrange dual problem can be given by
\vspace{-2mm}
\begin{alignat}{2}
\label{P3.2}
(\rm{P3.2}): \quad & \begin{array}{*{20}{c}}
\mathop {\max }\limits_{{ {\bm{\omega}} \in {\cal{P}}} } \mathop {\inf }\limits_\mu  L\left( {{\omega _1},...,{\omega _M}} \right)
\end{array}, & \\ 
\mbox{s.t.}\quad
& \mu \ge 0. & \nonumber 
\end{alignat} 
\par
At the optimal solution of (P3.2), $\frac{{\partial L\left( {{\omega _1},...,{\omega _M},{\mu }} \right)}}{{{\omega _m}}} = 0$. As $\omega_m > \tilde{\omega} _m$ and $\omega_{m'} > \tilde{\omega} _{m'}$, the following condition holds.
\vspace{-2mm}
\begin{equation}
\begin{aligned}
\frac{{\partial L\left( {{\omega _1},...,{\omega _M}} \right)}}{{{\omega _m}}} &=  \frac{{ - {W_{ - 1}}\left( {{x_m}  {e^{{x_m}}}} \right)}}{{\left( {{W_{ - 1}}\left( {{x_m}  {e^{{x_m}}}} \right) - {x_m}} \right)\left( {1 + {W_{ - 1}}\left( {{x_m}  {e^{{x_m}}}} \right)} \right)}} + \mu  \\
&= \frac{{ - {W_{ - 1}}\left( {{x_{m'}}  {e^{{x_{m'}}}}} \right)}}{{\left( {{W_{ - 1}}\left( {{x_{m'}}  {e^{{x_{m'}}}}} \right) - {x_{m'}}} \right)\left( {1 + {W_{ - 1}}\left( {{x_{m'}}  {e^{{x_{m'}}}}} \right)} \right)}} + \mu ,
\end{aligned}
\end{equation}
where $x_m = A_m \cdot \omega_{m}$, $x_{m'} = A_{m'} \cdot \omega_{m'}$, and $A_m = \frac{{C \ln 2  }}{{B  {\gamma _m}  {{\bar {E}}}}}$. As $\frac{{\partial L\left( {{\omega _1},...,{\omega _M}} \right)}}{{{\omega _m}}}$ increases monotonically with $\omega_{m}$, at the optimal solution of (P3.2), ${A_m}{\omega _m} = {A_{m'}}{\omega _{m'}}$. Hence, the following condition holds:
\begin{equation}
\frac{{{\omega _m}}}{{{\gamma _m}}} = \frac{{{\omega _{m'}}}}{{{\gamma _{m'}}}}.
\end{equation}

\footnotesize  	
\bibliography{mybibfile}
\bibliographystyle{IEEEtran}

\end{document}